# Resolving the era of river-forming climates on Mars using stratigraphic logs of river-deposit dimensions

Edwin Kite[1,2,3], Alan Howard[4], Antoine Lucas[5], Kevin Lewis[6].

1. Princeton University, Department of Geosciences.

2. Princeton University, Department of Astrophysical Sciences.

3. University of Chicago, Department of Geophysical Sciences, Hinds Building, 5734 S. Ellis Ave., Chicago IL 60637. (kite@uchicago.edu; t: 510-717-5205, f: 773-702-0207).

4. University of Virginia, Department of Environmental Sciences.

5. AIM CEA-Saclay, University Paris-Diderot.

6. Johns Hopkins University, Department of Earth and Planetary Sciences.

# 1. Introduction.

## 1.1. Fluvial signatures of climate events on Mars: clearer than on Earth.

# 2. Getting from stereopairs to stratigraphic logs.

# 3. Paleohydrology.

## 3.1. River-deposit scale varies systematically upsection.

## 3.2. Evidence for basin-wide paleodischarge variability.

## 3.3. Both short-term intermittency and long-term intermittency are required.

## 3.4. A tool to search for abrupt climate change on Mars.

# 4. Implications for river-forming paleo-climates on Mars.

## 4.1. Mechanisms for paleodischarge variability.

## 4.2. Global context.

# 5. Habitability, taphonomy, and organic matter preservation.

# 6. Conclusions.

**Supplementary Materials.**

    A. Supplementary Methods.

    B. Transect-by-transect stratigraphic logs.

# Abstract.

River deposits are one of the main lines of evidence that tell us that Mars once had a climate different from today, and so changes in river deposits with time tell us something about how Mars climate changed with time. In this study, we focus in on one sedimentary basin – Aeolis Dorsa – which contains an exceptionally high number of exceptionally well-preserved river deposits that appear to have formed over an interval of >0.5 Myr. We use changes in the river deposits' scale with stratigraphic elevation as a proxy for changes in river paleodischarge. Meander wavelengths tighten upwards and channel widths narrow upwards, and there is some evidence for a return to wide large-wavelength channels higher in the stratigraphy. Meander wavelength and channel width covary with stratigraphic elevation. The factor of 1.5-2 variations in paleochannel dimensions with stratigraphic elevation correspond to ~2.6-fold variability in bank-forming discharge (using standard wavelength-discharge scalings and width-discharge scalings). Taken together with evidence from a marker bed for discharge variability at ~10m stratigraphic distances, the variation in the scale of river deposits indicates that bank-forming discharge varied at both 10m stratigraphic ($10^2 - 10^6$ yr) and ~100 m stratigraphic ($10^3 - 10^9$ yr) scales. Because these variations are correlated across the basin, they record a change in basin-scale forcing, rather than smaller-scale internal feedbacks. Changing sediment input leading to a change in characteristic slopes and/or drainage area could be responsible, and another possibility is changing climate (±50 W/m$^2$ in peak energy available for snow/ice melt).

# 1. Introduction.

Many of the now-dry rivers on Mars were once fed by rain or by snow/ice melt, but physical models for producing that runoff vary widely (Malin et al. 2010, Mangold et al. 2004, Irwin et al. 2005a). Possible environmental scenarios range from intermittent <$10^2$ yr-duration volcanic- or impact-triggered transients to >$10^6$ yr duration humid greenhouse climates (Andrews-Hanna & Lewis 2011, Kite et al. 2013a, Mischna et al. 2013, Segura et al. 2013, Urata & Toon 2013, Tian et al. 2010, Halevy & Head 2014, Wordsworth et al. 2013, Kite et al. 2014, Ramirez et al. 2014). These diverse possibilities have very different implications for the duration, spatial patchiness, and intermittency of the wettest (and presumably most habitable) past climates on Mars. As a set, these models represent an embarrassment of riches for the Mars research community, and paleo-environmental proxies (ideally, time series) are sorely needed to discriminate between the models. In principle, discriminating between the models using the fluvial record should be possible, because runoff intensity, duration and especially intermittency control sediment transport (e.g. Devauchelle et al. 2012, Morgan et al. 2014, Williams & Weitz 2014, Erkeling et al. 2012, Kereszturi 2014, Kleinhans et al. 2010, Barnhart et al. 2009, Howard 2007). For example, Barnhart et al. (2009) and Hoke et al. (2011) both use geomorphic evidence for prolonged sediment transport to argue forcefully that impact-generated hypotheses for Early Mars runoff cannot be sustained.

However, geomorphology usually provides time-integrated runoff constraints, whereas constraining climate change requires time-resolved constraints; few locations on Mars show evidence for more than one river-forming episode; and correlation between those locations has to rely on crater counting, which (for this application) suffers from small-number statistics, cryptic resurfacing, target strength effects, confusion between primary and secondary craters, and inter-analyst variability (e.g. Dundas et al. 2010, Smith et al. 2008, Warner et al. 2014, Robbins et al. 2014). These problems have limited the application of dry-river evidence to constrain climate change during the era of river-forming climates on Mars.

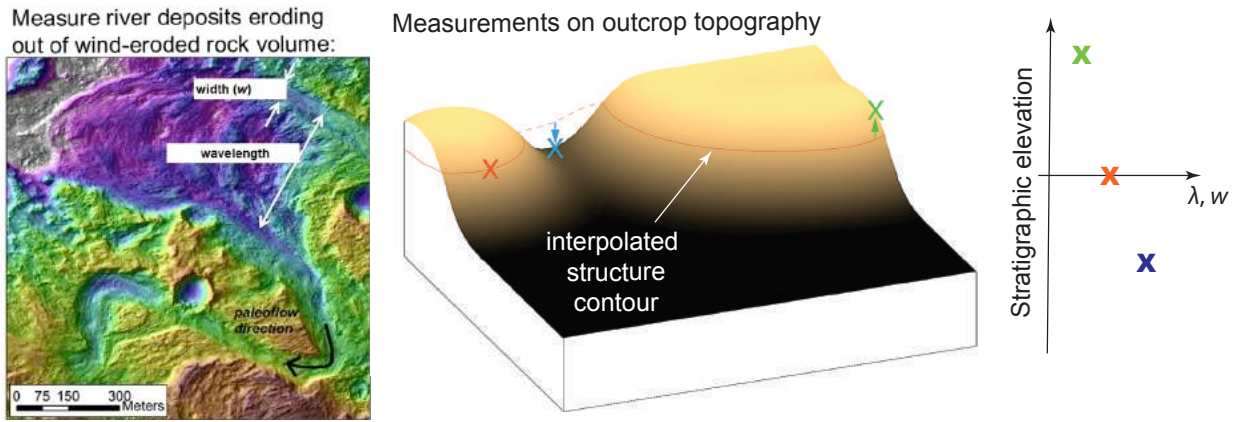

**Figure 1**. Graphical abstract of this study. We measure the width ($w$) and wavelength ($\lambda$) of Mars river-deposits eroding out of the rock, assign stratigraphic elevations $z_s$ to each measurement, and convert these to estimates of paleo-discharge $Q(z_s)$. Depth of chute cutoff (in left panel) is 2 m.

Aeolis Dorsa (1°S-8° S, 149°E-156° E) – a wind-exhumed sedimentary basin 10° E of Gale crater that contains an exceptionally high number of exceptionally well-preserved river deposits – gets around these problems. At Aeolis Dorsa, basin-scale mapping distinguishes $10^2$ m-thick river-deposit-hosting units, which collectively provide time-resolved climate constraints (river valleys provide time-integrated constraints) (Kite et al., submitted). We can put these deposits in time order using crosscutting relationships, which lack the ambiguity of crater counts. Paleodischarge can be estimated from meander wavelengths and channel widths (Burr et al. 2010). Therefore, Aeolis Dorsa contains a stratigraphic record of climate-driven surface runoff on Mars (Burr et al. 2010; Fairén et al., 2013; Kite et al. submitted).

To constrain paleodischarge versus time, in this study we measure how river-deposit dimensions vary with stratigraphic elevation (Hajek & Wolinsky 2012). The key results are set out in Table 1. We review terrestrial background and geologic context in §1.1. We introduce our method (Fig. 1) in §2; this method is generally applicable to stratigraphic logging from stereopairs (not just logs of river-deposit dimensions). We report our dataset, describe our paleodischarge interpretation, and discuss the implications for fluvial intermittency and abrupt climate change in §3. We discuss implications for paleodischarge variability in §4, assess the science merit of landing at Aeolis Dorsa in §5, and conclude in §6.

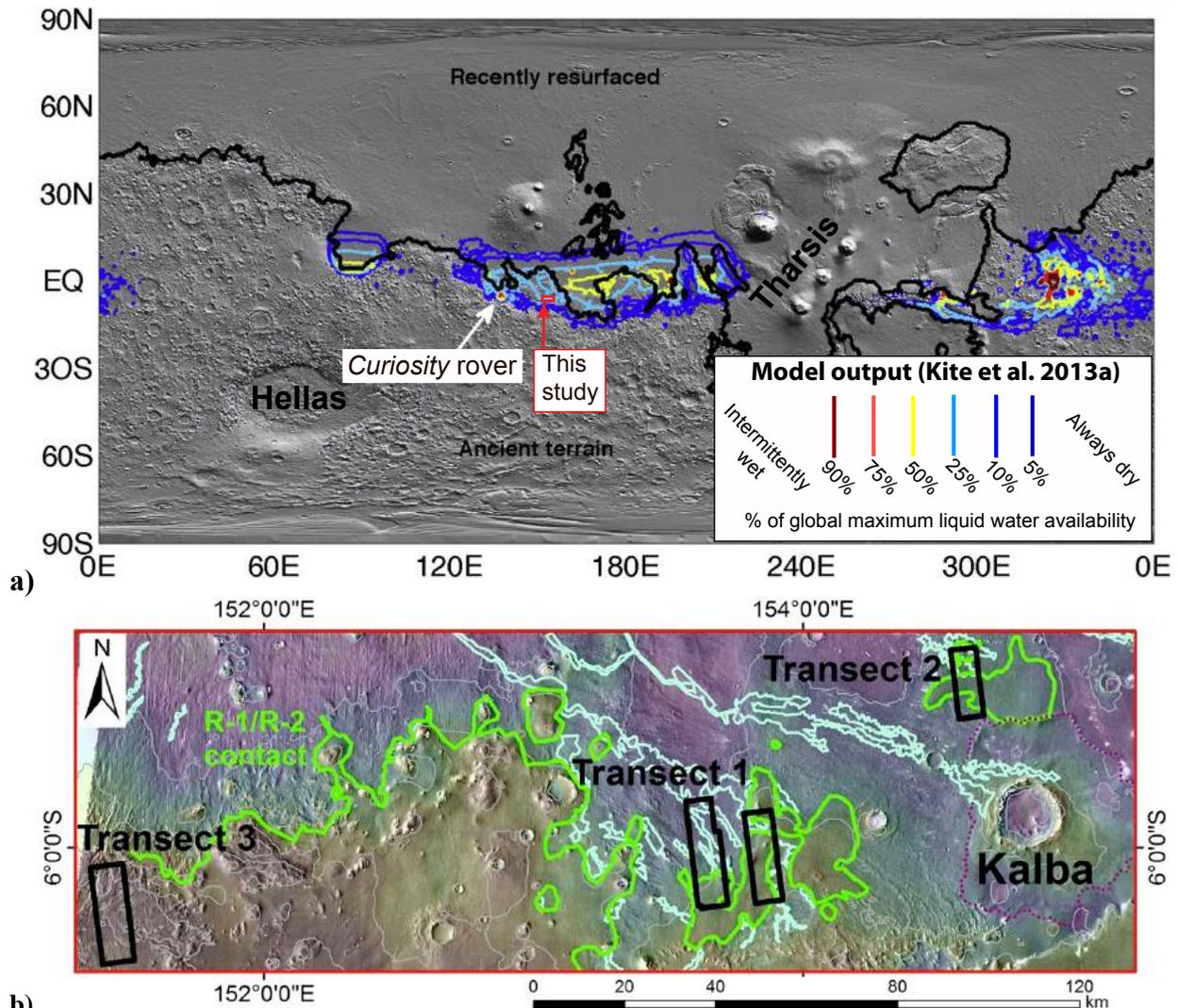

**Figure 2. a)** Locator map for our study area (red rectangle). Background is shaded relief Mars Orbiter Laser Altimeter (MOLA) topography, illuminated from top right. Colored contours show output from a seasonal melting model (Kite et al. 2013a: relative frequency of years with seasonal surface liquid water, average of 88 different orbitally-integrated thin-atmosphere simulations). The black line marks the border of the area of recently-resurfaced terrain, and approximately corresponds to the hemispheric dichotomy. This figure is modified from Fig. 16d in Kite et al. 2013a. **b)** Zooming in to our study area, showing locations of transects (black rectangles; detail shown in Supplementary Materials Section A2). Transect 1 consists of two nearby non-contiguous areas. Green line shows trace of R-1/R-2 contact ($z_s$ = 0 m). R-2 is above the contact (orange/red tints), and R-1 is below the contact (purple/blue tints). Grey contours are MOLA topography, at 200 m intervals. Cyan outlines correspond to large meander belts, which disappear beneath and reappear from underneath smooth dome-shaped outcrops of R-2. Purple dashed line outlines an old crater.

## 1.1. Fluvial signatures of climate events on Mars: clearer than Earth.

Fluvial sediments record climate events in Earth history through changes in river-deposit dimensions, channel-deposit proportions, and fluvial styles (e.g. Foreman et al. 2012, Macklin et al. 2012, Amundson et al. 2012, Schmitz & Pujalte 2007).

Dry rivers are most useful in reconstructing climate change when the record is preserved as deposits (as at Aeolis Dorsa) and when it is uncontaminated by large-amplitude externally-driven tectonic uplift (as at Aeolis Dorsa). Mars lacks plate tectonics and has been tectonically quiescent for >3 Ga (Golombek & Phillips 2012), making the fluvial record of climate change clearer than on Earth where the strong effects of base-level fluctuations and synfluvial tectonics complicate interpretation of the fluvial record in terms of climate change (e.g. Blum & Törnqvist 2000). Reviewing Earth work, Whittaker (2012) states "If topography forms a non-unique or difficult-to-decode record of past climate […], it is likely that the sedimentary record, if and where complete, forms the best archive of landscape response to past climate."

The sedimentary record in Aeolis Dorsa is >3 km thick (Kite et al., submitted), and this study focuses on ~300m of stratigraphy bracketing the contact (green line in Fig. 2b) between two river-deposit-containing units that show dramatically different erosional expression. These units are termed R-1 (low-standing) and R-2 (high-standing). In both units, river deposits outcrop in plan view. Preservation quality is better than for any on-land plan-view outcrop on Earth (Fig. 1), and resembles high-resolution 3D seismic data (e.g. Reijenstein et al. 2011, Hubbard et al. 2011). Crater-counts and stratigraphic analyses (Zimbelman & Scheidt 2012) correlate the older rivers that are currently exposed in outcrop to either the central Hesperian or to the Noachian-Hesperian transition (thought to have been a habitability optimum; e.g., Irwin et al. 2005b).

## 2. Getting from stereopairs to stratigraphic logs.

We seek constraints on river discharge $Q$ as a function of stratigraphic elevation $z_s$ from deposits containing proxy data (such as channel width $w$ and meander wavelength $\lambda$) as a function of 3D spatial position (latitude, longitude, and elevation). Therefore, we must (1) convert 3D spatial position to $z_s$, and (2) use a transfer function to convert proxy-data averages to $Q(z_s)$ (Hajek & Wolinsky 2012, Irwin 2007, Burr et al. 2010, Williams et al. 2009). (The following is a brief discussion; for more details, see the Supplementary Materials.)

The stratigraphic interval of interest is exposed over an area ($A_b$) of $\sim 2 \times 10^4$ km². Assuming a drainage density of 1/km (mapped channels), a stratigraphic separation between channels of 10 m, and an averaged preserved thickness of 300 m, the length of preserved channel deposits in $A_b$ is comparable to that of all channels previously mapped on Mars ($\sim 8 \times 10^5$ km; Hynek et al. 2010). In terrestrial fieldwork we would tackle such data overload by defining transects and measuring a stratigraphic section along each transect. The analogous procedure to defining a transect for Mars orbiter data is to select high-resolution stereopairs (each stereopair covers an area of ~100 km² $<< A_b$), each chosen to span a wide stratigraphic range. Each stereopair is converted to a digital terrain model (DTM) and accompanying orthorectified images using BAE Systems SOCET SET software (Kite et al. 2014). We use 5 stereo DTMs in total, defining three transects that collectively span the area where R-2 is exposed (Fig. 1; Supplementary Table 1; Figs. S1-S6). Three of the DTMs come from an area of exceptionally good preservation and dense High-Resolution Imaging Science Experiment (HiRISE) coverage near 153.6°E 6°S, which we term Transect 1.

For each transect, we map out $z_s$ by subtracting an interpolated R-1/R-2 contact surface from the topography, so that the R-1/R-2 contact has $z_s = 0$ m. First we define the contact surface between R-1 and R-2 by fitting a surface to points where the large meander-belts characteristic of R-1 were drowned or blanketed by smoothly-eroding R-2 deposits. (There is evidence for erosion between R-1 and R-2, so the contact is probably time-transgressive in detail; Supplementary Materials). Second-order polynomial fitting is used; similar results were obtained using other deterministic interpolation methods. The layers are nearly flat (typically <1 °), so the elevation difference is also the stratigraphic offset.

Next, we pick channel centerlines, channel banks, areas of channel deposits, and areas of evidence for lateral migration of channels during aggradation. Every vertex from these picks is tagged with a stratigraphic elevation $z_s$.

To get $\lambda$, each channel centerline is picked 3 times by hand. We then convert into coordinates of along-centerline distance $s$ and direction $\theta$ and identify half-meanders using inflection points (zero-crossings of $\partial\theta/\partial s$) (Fig. 3) (Howard & Hemberger 1991). Sinuosity is defined as along-centerline distance between inflection points divided by straight-line distance between inflection points. Half-meanders with sinuosity <1.1 are rejected. For half-meanders with sinuosity ≥1.1, we take $\lambda$ as twice the distance between inflection points.

To get $w$, we interpolate equally-spaced points along each bank, and then find the closest distance from each point to the opposing bank. $w$ is then defined as the median bank-to-bank distance.

We need error bars to avoid overfitting models to data. There are four principal contributors to error: $z_s$ error, measurement error, preservation bias, and sampling error. $z_s$ error is quantified using the root-mean-square error on the interpolated contact surface. The stratigraphic RMS error is 20 m for Transect 1; 5 m for Transect 2; and 67 m for Transect 3 (Supplementary Methods). Measurement error is quantified (for $\lambda$) using the variance of $\lambda$ extracted from repeated, independent traces on the same channel centerlines (Fig. 3) and (for $w$) by using the variance of bank-to-bank widths measured at different locations along the channel trace. Preservation bias refers to erosional narrowing (or widening) of $w$ (Williams et al. 2009). In practice, we found preservation bias to be minor (§3.2-§3.3, Fig. S7). Sampling error is quantified by bootstrapping. Bootstrapping accounts for the fact that we can only measure the subset of river deposits that outcrop in our DTM areas, which is a chance sampling from river-deposits that are dispersed within a three-dimensional rock volume. When applying the bootstrap, we must combine non-independent measurements, otherwise the error bars from the bootstrap will be too small. We therefore treat data points collected from the same channel thread as a single combined measurement. The bootstrap error bars shown in Fig. 4 and Fig. S2 take into account $z_s$ error and measurement error as well as sampling errors. We log-transform all data before calculating errors and averages, because we are interested in relative changes in scale parameters.

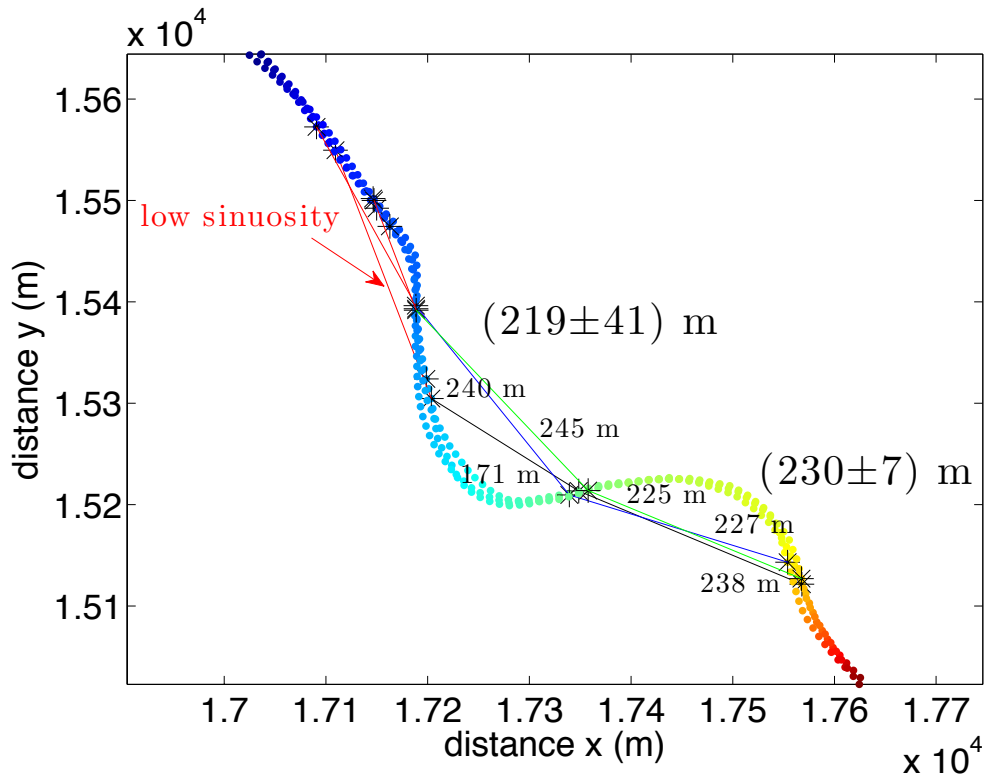

**Figure 3.** Wavelength (λ) extraction from repeated ArcGIS traces. Points are uniformly interpolated along three independent centerline traces (colored disks). Disk color corresponds to the modern topographic range – standard deviation of 4m in this case. Lines link inflection points (asterisks) that are obtained from each centerline trace. The low-sinuosity half-meander (arrowed) is excluded from the analysis.

## 3. Paleohydrology.

### 3.1 River-deposit scale varies systematically upsection.

Wavelength $\lambda$ tightens upwards across the contact ($z_s = 0$ m in Fig. 4a). In the best-exposed interval (-100 m < $z_s$ < 130 m), small meanders are rare/absent below 0 m, and common above 0

m. Visual inspection of the data suggests that the most natural break in the data (i.e. the paleohydrologically-inferred contact) is 20m below the geologically inferred contact (the two-sample Kolmogorov-Smirnov test statistic is minimized for a breakpoint at $z_s = -16.5$ m for $\lambda$). From $10^5$ Monte Carlo bootstrap trials, in no cases did $\widetilde{\lambda}(z_s > -20$ m$)$ exceed $\widetilde{\lambda}(z_s < -20$ m$)$ (where $\widetilde{\lambda}$ is the sample median). The same result was obtained for a break-point at $z_s = 0$m. The trends are consistent between Transect 1 and Transect 2 (Figs. S3, S5-S6).

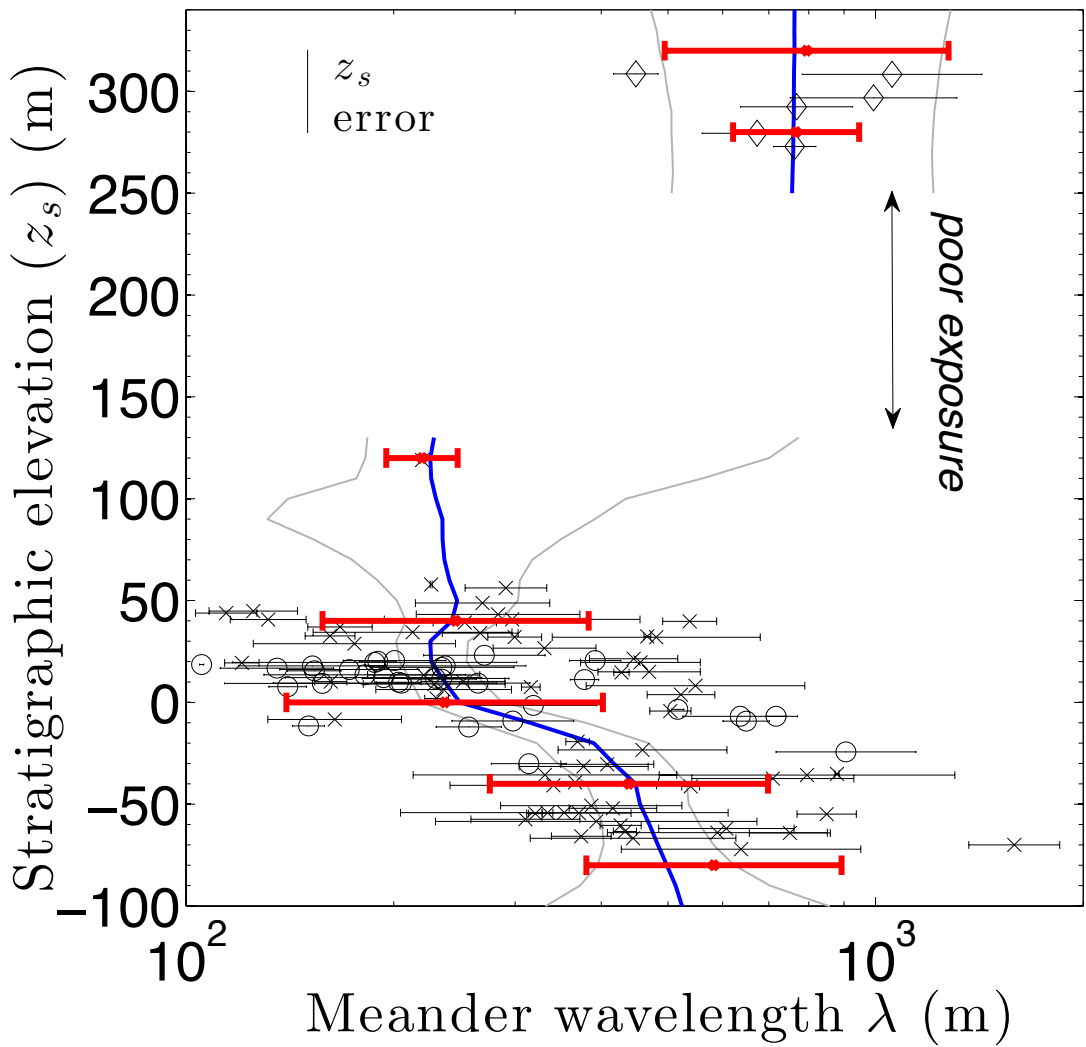

a)

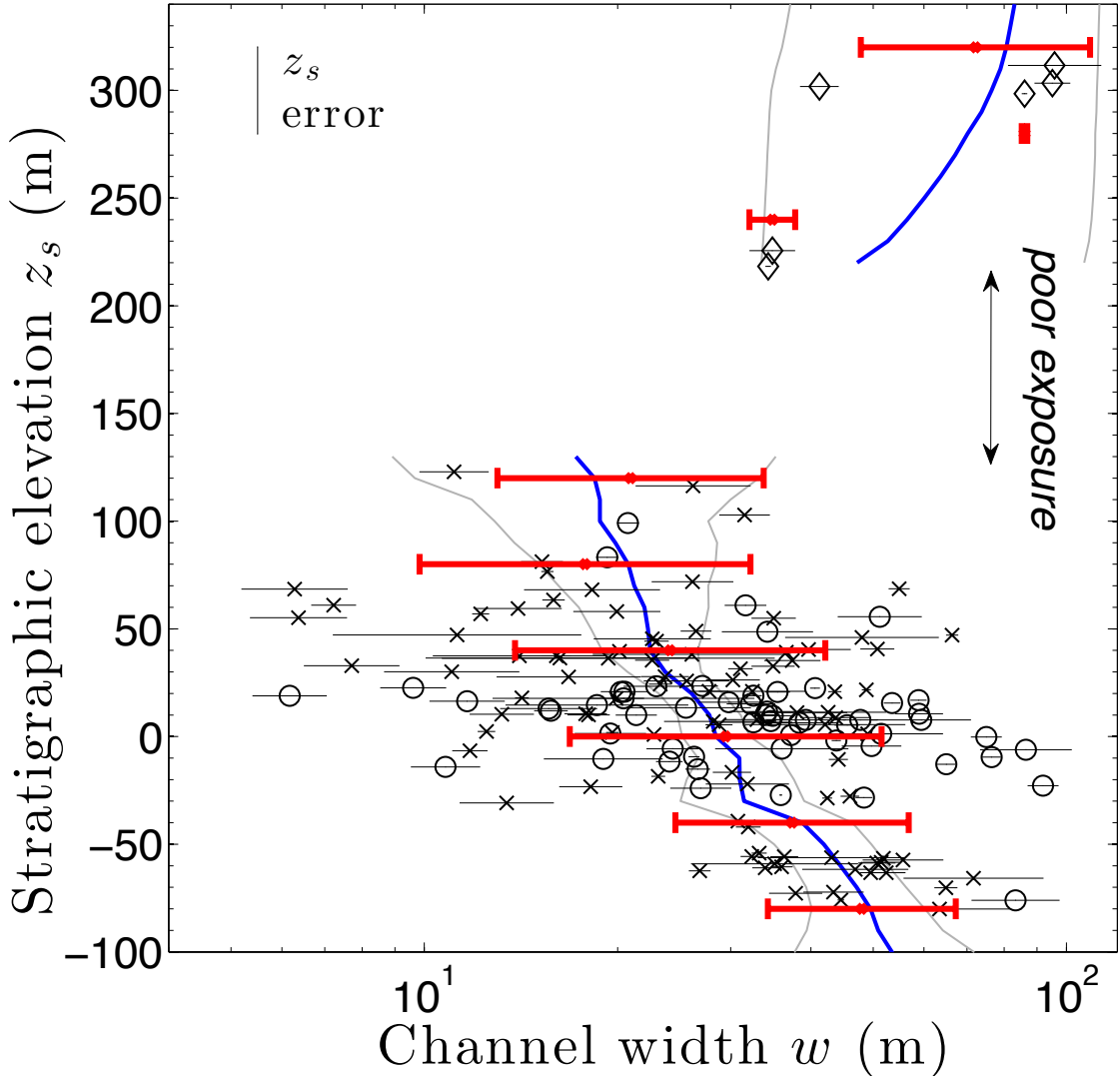

b)

**Figure 4**. Width ($w$) and wavelength ($\lambda$) results. The black symbols show the measurements; crosses × for Transect 1, circles ○ for Transect 2, and diamonds ◇ for Transect 3. The error bars in $w$ and $\lambda$ differ for each measurement and are shown individually. The error bars in $z_s$ shown in the top left of each plot correspond to the median $z_s$ errors. The red error bars show the mean and standard deviation of the data in 40 m-wide $z_s$ bins (40 m is approximately ~2 × the median $z_s$ error). The blue line shows a running estimate of the average (for a 40 m $z_s$ window), from a bootstrap. The gray lines correspond to the 2-σ errors on the running average from the same bootstrap, which are equivalent to the (2.1% - 97.9%) bounds from the bootstrap. Gaps in the gray and blue lines show range of $z_s$ with few or no data. Fig. S2 shows the $z_s$ = -100 m to $z_s$ = + 100 m interval in more detail.

$w$ also narrows upwards across the contact ($z_s$ = 0m in Fig. 4b). In the best-exposed interval (-100 m < $z_s$ < 130 m), narrow channels are rare/absent below 0 m, and common above 0 m.

Similar to $\lambda$, visual inspection of the data suggests that the most natural break in the data (i.e. the paleohydrologically-inferred contact) is $z_s \sim -20$m below the geologically inferred contact. (The two-sample K-S test statistic is minimized for a breakpoint at $z_s = +16$m, which reflects the greater density of data points at $z_s = +20$m to $+50$m). From $10^5$ Monte Carlo bootstrap trials, in only 1 case the median $w$ above -20m exceeded the median $w$ below -20m. For a break-point at $z_s = 0$m, the same test showed that only 9 cases out of $10^5$ with $\widetilde{w}(z_s < -20\text{ m}) < \widetilde{w}(z_s > -20\text{ m})$. The trends are consistent between Transect 1 and Transect 2 (Figs. S5-S6).

Width $w$ and wavelength $\lambda$ are poorly constrained for $z_s > 130$ m. There is some evidence for a return to bigger rivers at higher $z_s$. While there are relatively few data points here (all from Transect 3; diamonds), the trend to higher $\lambda$ and $w$ in Transect 3 is statistically significant and visually obvious in the DTMs. However, assigning Transect 3 data points to $z_s > 130$m is only correct if we assume that the topographic offset between the large rivers in Transect 3 and the large rivers in Transect 1 (Fig. 2b) corresponds to a stratigraphic offset. This assumption could be wrong if the river-deposits measured in Transect 3 are separated from the river-deposits measured in Transects 1 and 2 by a drainage divide (Kite et al., submitted). Drainage divides can separate nearby rivers formed at the same time and with large topographic offsets (e.g., Eastern Continental Drainage Divide, Blue Ridge Scarp, North Carolina, USA). The topographically-offset Transect 3 deposits could be stratigraphically equivalent to the big river-deposits at $z_s \approx -50$m in Fig. 4 if they are separated from them by a drainage divide. The drainage divide possibility could be tested in future by using meander-migration directions to test for paleoflow direction.

Sinuosity decreases upsection (Fig. S4) but stays low for Transect 3.

a)

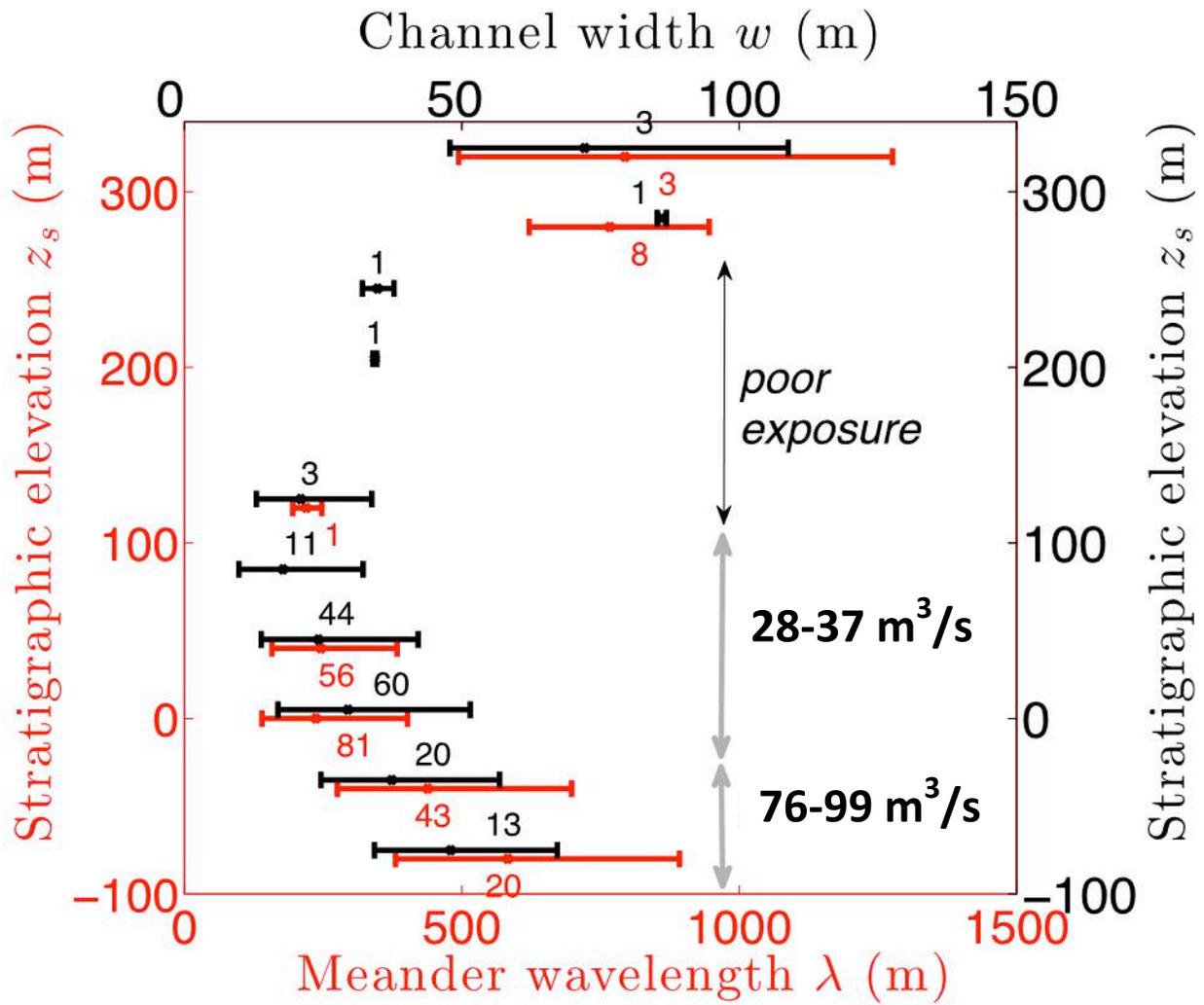

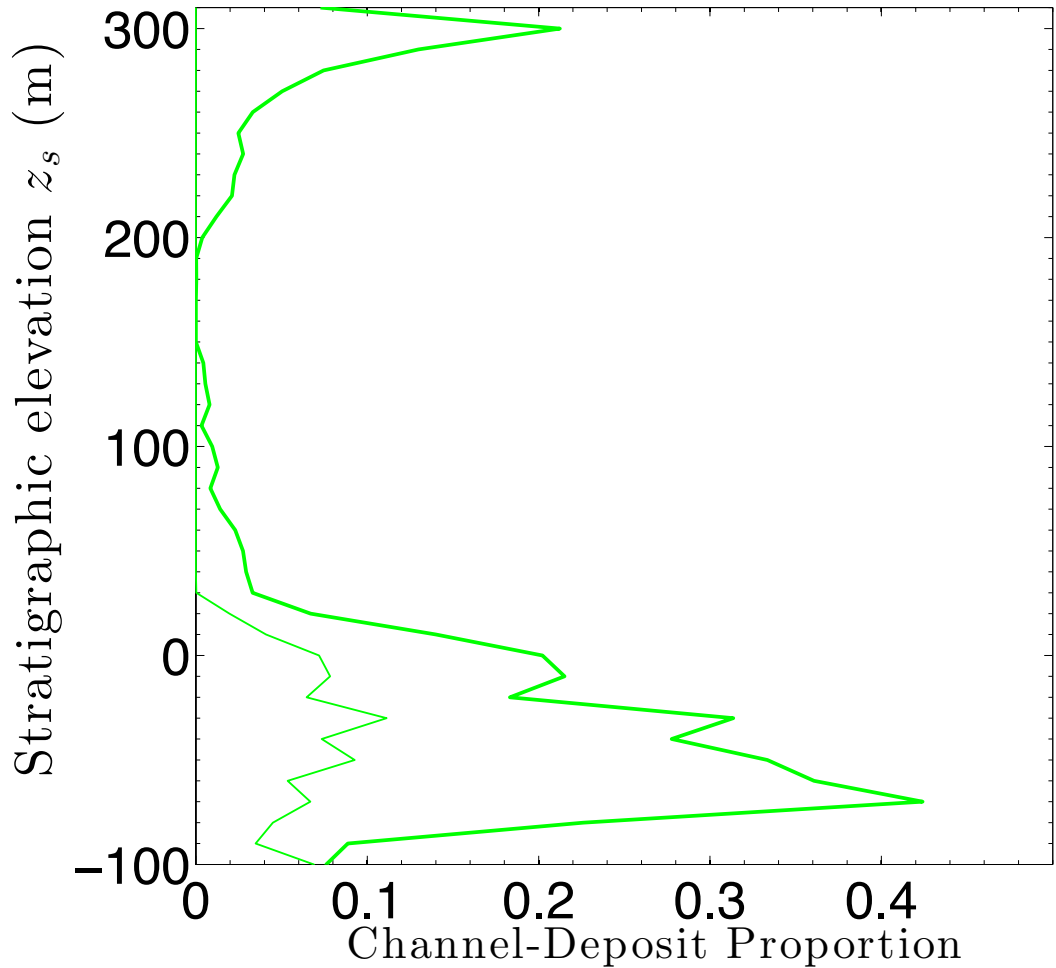

**b)**

**Figure 5.** Summary of results, combining all transects. **a)** Comparison of width (*w*) & wavelength (*λ*) results. Error bars show 1σ spread of measurements (averaged in 40 m $z_s$ bins). Numbers next to error bars correspond to the number of measurements. **b)** Areal fractions of channel deposits (thick line) and of deposits showing evidence for lateral accretion (thin line). $z_s$ bin size = 10 m.

Cross-checking of widths and wavelengths suggests that our measurements of *w* have comparable reliability to our measurements of *λ* (Fig. 5a). This is surprising, because it is much easier for erosion to narrow inverted channels (Williams et al. 2009) and widen valleys than to significantly modify $\lambda$. This consistency reflects good preservation of fluvial form.[1]

To support our scale measurements, we record the channel-deposit outcrop proportion and the fractional exposure of lateral accretion deposits versus $z_s$ (Fig. 5b). Channel-deposit proportion is high where river-deposit dimensions are large. Evidence for lateral migration of channels during aggradation is frequent where rivers are large, and very uncommon where rivers are small. Results are very similar in both Transect 1 and Transect 2 (not shown).

All these metrics are self-consistent, as expected: larger width rivers should make proportionally larger meander bends (with $w:\lambda \approx 10$, as observed), and should also migrate laterally at faster rates.

## 3.2. Evidence for basin-wide paleo-discharge variability.

The decrease in both $\tilde{\lambda}$ and $\tilde{w}$ that is apparent in both Transect 1 and Transect 2 (Fig. 4) is representative of a basin-wide change in channel-deposit dimensions. The reduction in $\tilde{\lambda}$ and $\tilde{w}$ upsection is visually obvious in the transect DTMs (Fig. S1) and also in ConTeXt Camera (CTX) mosaics spanning the full area of the R-1/R-2 contact. Basinwide, R-1 erodes to form yardangs and is very rough at km scales, not retaining recognizable craters; river deposits in R-1 formed large meanders and broad meander belts. Basinwide, R-2 is topographically high-standing, smoothly eroding, associated with numerous aeolian bedforms, and retains many craters. River deposits in R-2 left narrow channel deposits and formed tight meanders. The basin-wide spatial scale and ~100m stratigraphic scale of the changes we report in this study probably puts them beyond the grasp of shredding of environmental signals by nonlinear sediment transport (Jerolmack & Paola 2010, Sadler & Jerolmack, 2014).

---

[1] In spite of our expectation that preserved $w$ would be a poor paleodischarge proxy, we find that:- $\tilde{w}:\tilde{\lambda}$ is consistent with measurements of Earth rivers (Bridge 2003); inverted-channel $w$ is not significantly narrower than negative-relief $w$ (Fig. S6); the fractional standard deviations of $w$ for a given $z_s$ are comparable to those for $\lambda$ (Fig. 4); for those $\lambda$ for which $w$ was also measured, most have a $w:\lambda$ ratio within range of terrestrial measurements (Fig. S7); and, again for $\{\lambda, w\}$ pairs, there is no evidence for $w:\lambda$ ratio changing with $z_s$. With hindsight, this consistency is likely aided by intentional selection of only the best-preserved stretches of channel in each DTM. From that subset of measured channels, those that showed signs of relatively more severe erosion were rejected as "candidates." This left a fairly restricted set of relatively high-quality measurements. An unrestricted sampling of all channel stretches without consideration of preservation quality would provide less reliable width measurements.

Motivated by the coherence of our paleohydrologic proxies – between metrics and between transects (Figs. S3-S6) – we conclude that R-2 records a distinct episode of runoff from R-1. We now turn to quantifying the change in discharge across the contact by using $\lambda$ and $w$ as proxies for $Q$.

For rivers on Earth, bank-full discharge $Q$ scales approximately as $w^2$ and as $\lambda^2$. The physical basis for this increase (Finnegan et al. 2005) is that channel depth $h$ tends to increase in proportion with $w$, whereas water velocity $u$ increases only sluggishly with $w$. Since $Q = whu$ (by continuity), $Q \propto w^{\sim 2}$. Theory suggests that initially-straight rivers are most vulnerable to sinuous instabilities with $\lambda \sim 10\ w$ (bar-bend theory) (Blondeaux & Seminara 1995), and that as the meander amplitude grows, the wavelength of the initial instability is frozen-in to the growing meander (Bridge 2003). Therefore $\lambda \propto w$ (Bridge 2003).

We adopt the Eaton (2013) scaling for $w$ and the Burr et al. (2010) scaling for $\lambda$. Rivers on Mars are expected to flow slower than rivers on Earth because Martian gravity is lower. Theoretically, $w \propto g^{-0.2335}$ (Parker et al. 2007), so we divide the widths by (Mars gravity / Earth gravity) $^{-0.2335}$ = 1.257. Therefore:

(1) $\quad Q = ( w / (1.257 \times 3.35 ) )^{1.8656}$

(2) $\quad Q = 0.011 ( \lambda / 1.267)^{1.54}$

We report 'nominal' bank-full discharges in Table 1.[2,3]

|  | $\widetilde{\lambda}$(m) | Bankfull $Q$ from $\widetilde{\lambda}$ (m$^3$ s$^{-1}$) | $\widetilde{w}$(m) | Bankfull $Q$ from $\widetilde{w}$ (m$^3$ s$^{-1}$) |
|---|---|---|---|---|
| Interpolated $z_s$ | | | | |
| 100m > $z_s$ > -20m | (246 ± 15) m | (37 ± 3) m$^3$ s$^{-1}$ | (25 ± 1.5) m | (28 ± 3) m$^3$ s$^{-1}$ |
| $z_s$ < -20m | (466 ± 35) m | (99 ± 12) m$^3$ s$^{-1}$ | (43 ± 3.7) m | (76 ± 12) m$^3$ s$^{-1}$ |
| Interpolated $z_s$ | | | | |
| 100m > $z_s$ > 0m | (229 ± 14) m | (33 ± 3) m$^3$ s$^{-1}$ | (25 ± 1.5) m | (28 ± 3) m$^3$ s$^{-1}$ |
| $z_s$ < 0m | (432 ± 31) m | (89 ± 10) m$^3$ s$^{-1}$ | (37 ± 3.2) m | (58 ± 9) m$^3$ s$^{-1}$ |
| Unit assigned from mapping (no interpolation) | | | | |
| R-2 | (219 ± 14) m | (31 ± 3) m$^3$ s$^{-1}$ | (22 ± 1.5) m | (22 ± 3) m$^3$ s$^{-1}$ |
| R-1 | (406 ± 29) m | (80 ± 9) m$^3$ s$^{-1}$ | (37 ± 3.0) m | (58 ± 9) m$^3$ s$^{-1}$ |
| Transect 3 | (764 ± 82) m | (213 ± 35) m$^3$ s$^{-1}$ | (61 ± 13) m | (147 ± 58) m$^3$ s$^{-1}$ |

**Table 1.** Paleodischarge $Q$ versus stratigraphic elevation $z_s$ from transects in S. Aeolis Dorsa.

Paleo-discharges inferred for unit R-1 are higher than those for R-2 by ~260%. This does not depend on how the breakpoint is defined (Table 1). $Q_\lambda$ is higher than $Q_w$, but within the standard error of the terrestrial regressions (41% standard error for $\lambda$, and 29% for $w$) (Burr et al. 2010, Eaton 2013). This difference might be caused by erosional narrowing of channel widths.

---

[2] More sophisticated approaches to estimating paleo-discharge require slope and/or grain-size information. Unfortunately, in southern Aeolis Dorsa, the present-day slopes of the river deposits are unlikely to be a safe guide to the slopes when the rivers were flowing (Lefort et al. 2012), and clasts are not seen in HiRISE images.

[3] $w$-$Q$ relationships in ice-floored channels are poorly constrained (e.g. McNamara & Kane 2009). Strong banks are needed for meandering; on Mars, ice, salt or clay could firm up banks (Matsubara et al. submitted).

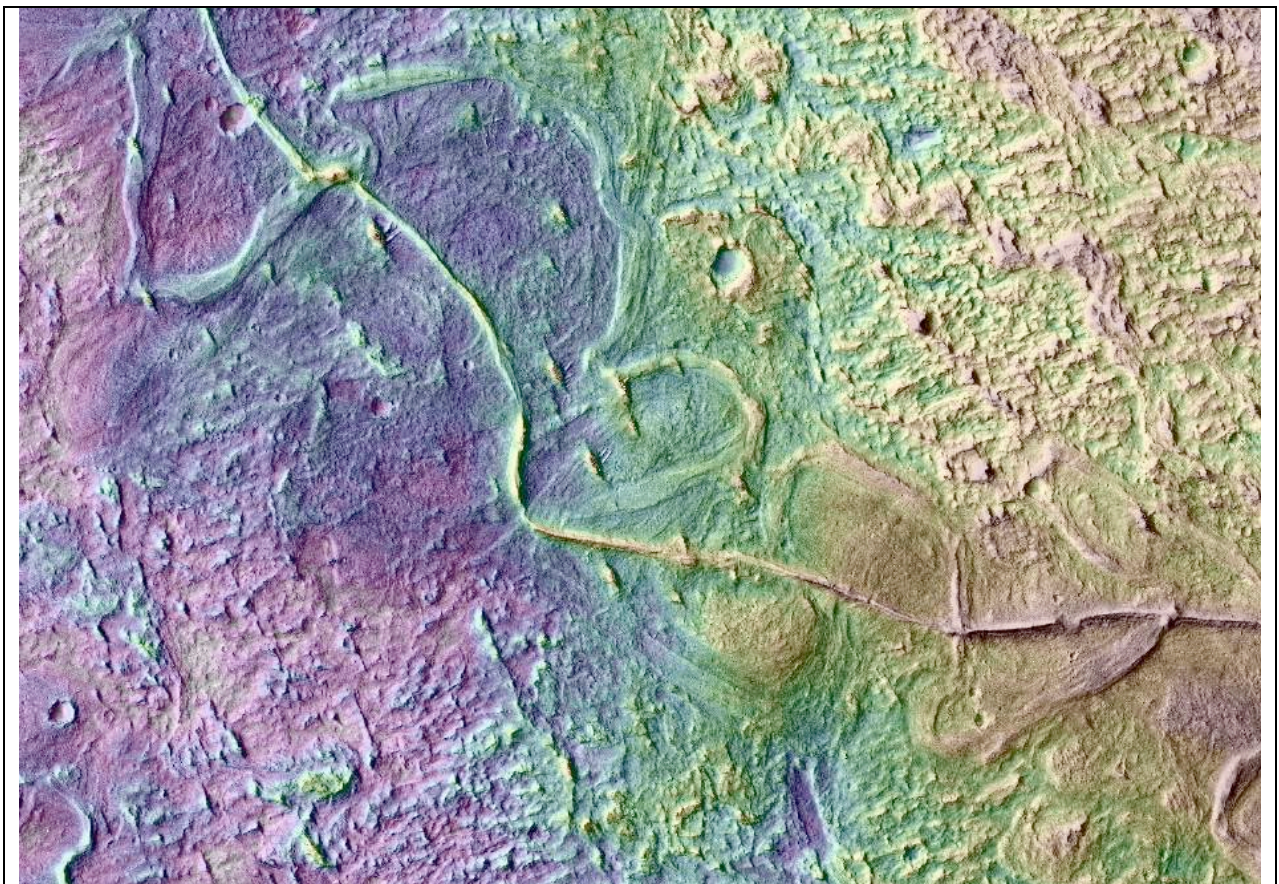

**Figure 6.** Evidence for intermittency at approximately 10m stratigraphic scale from DTM constructed from PSP_006683_1740 and PSP_010322_1740. Image is 4.2 km across. The less-sinuous inverted channel is ~10 m above the more-sinuous early meander belt. The full range of topography in the DTM is ~30 m (which may include postdepositional tilting; Lefort et al., 2012). Earlier channel generations which were crosscut by the early meander belt are visible in top right (pooly-preserved sinuous ridges).

In addition to the large-scale changes shown in Table 1, our DTMs record paleodischarge variability at $z_s$ scales too fine to be resolved in our stratigraphic logs. For example, Fig. 6 shows a low-sinuosity inverted channel without cutoffs that follows the trend of a very sinuous meander-belt deposit with numerous cutoffs (Burr et al. 2010, Jacobsen & Burr 2013, Cardenas

& Mohrig 2014). The inverted channel is ~10m above the meander belt. This sequence is widespread in southeast Aeolis Dorsa (e.g. HiRISE images ESP_036378_1740 and PSP_008621_1750). We interpret this pattern as a stratigraphic marker of a basinwide event (rather than the result of uncorrelated avulsions), because the pattern is never seen at more than 1 stratigraphic level locally, has a common erosional expression basinwide, joins at confluences, and is only found just below the R-1/R-2 contact. The following cut-and-fill sequence may explain these observations:

- Rivers transported sediment for long enough to develop high-sinuosity channels and a broad meander belt with cutoffs.

- Fluvial sediment transport ceased, or was diverted. The meander-belts were then covered by ~10m of now-removed cover material, which infilled the channels but failed to completely mute the paleo-valleys containing the meander-belts.

- The channel-belt cover was re-incised by rivers that followed the trend of the partly-infilled valleys.

- Fluvial sediment transport shut down again – after a shorter period of fluvial activity than for the first-generation channels – and the less-sinuous valley was itself infilled.

- Late-stage erosion removed the cover material, causing pronounced topographic inversion of the less-sinuous channel and partial topographic inversion of the more-sinuous channel. Preferential removal of the cover material by wind erosion is consistent with the idea that the cover material was wind-blown: in Aeolis Dorsa, areas that were once wet (e.g., meander belts, channels) are preferentially preserved against modern wind erosion.

The inferred switching from fluvial sediment transport, to infilling/mantling, followed by fluvial re-incision, is consistent with a wet-dry-wet sequence (Metz et al. 2009, Williams & Weitz 2014). We also find cut-and-fill cycles in R-2, with stratigraphic amplitude ~3m. However, there is no requirement from our DTMs that *peak* flows during the event that formed the less-sinuous channels were less than *peak* flows during the event that formed the more-sinuous underlying meander belt. We hand-tagged $\lambda$ and $w$ measurements in our DTMs corresponding to the two layers (which are visually distinctive and easy to identify; Fig. 6). We find $\lambda$ = 588 (+631/–304) m ($n$ = 8) for the more-sinuous meander belt and $\lambda$ = 421 (+191/–132) m ($n$ = 13) for the less-sinuous inverted channel. These results overlap, although the more-sinuous meander belt does have some tight, short-wavelength meanders which are not found in the less-sinuous inverted channel. We find $w$ = 46 (+18/–13) m ($n$ = 19) in the more sinuous meander belt and $w$ = 39 (+12/–9) m ($n$ = 11) for the less sinuous member. These results are not significantly different.

### 3.3. Both short-term intermittency and long-term intermittency are required.

There is evidence for intermittency at four different stratigraphic scales in Aeolis Dorsa. (1) Erosion-deposition alternations suggest wet-dry alternations in Aeolis Dorsa at the stratigraphic scale of hundreds of meters to kilometers (Kite et al., submitted). Aeolis Dorsa's > 3 km of stratigraphy consists of four rock packages bounded by unconformities. The top three rock packages probably required near-surface liquid water to make rivers, make alluvial fans, and cement layers. The unconformities show erosion and interbedded craters suggesting long time gaps (Kite et al., submitted). Aeolian erosion during these time gaps is suggested by smooth deflation at the unconformities, and if surface liquid water been available it would have suppressed aeolian erosion by trapping the clasts needed for wind-induced saltation abrasion. (2) At 100 m stratigraphic scale, our logging suggests variations in discharge (Figs. 4-5). (3) At 10 m stratigraphic scale, we interpret the stratigraphy of the marker bed (Fig. 6) to suggest wet-dry-wet alternations. (4) At 1 m stratigraphic scale, chute cutoffs indicate discharge variability, and it is also tempting to interpret the 2-3m banding in the lateral-accretion deposits as the result of annual floods.

What was the timescale for these processes? The high density of embedded impact craters suggests a depositional interval of >(4-20) Myr (Kite et al. 2013b). This method does not constrain whether deposition was steady or pulsed. Sediment transport calculations provide an estimate of the intermittency. To re-incise a fresh channel such as the low-sinuosity inverted channel in Fig. 6, the volume of the channel must be transported downstream as sediment (Church 2006). Therefore $\tau = L_c\, w\, h\, /\, (\, Q\, f_{sed}\, )$, where $\tau$ is timescale, $L_c$ is channel length, $h$ is channel depth, $Q$ is water discharge, and $f_{sed} \ll 1$ is sediment fraction. Setting $L_c \approx 10^2$ km (Fig. 2b), $w \approx 40$ m (Table 1), $h \approx w / 57$ (Hajek & Wolinsky 2012), $Q \approx 80$ m$^3$/s, and $f_{sed} = 1\text{-}2 \times 10^{-4}$ (Palucis et al. 2014), we obtain $\tau \approx 0.15\text{-}0.3$ yr. If the entire volume of unit R-1 was transported by rivers, $\tau = \Delta z_s\, A_b\, /\, (\, N_r\, Q\, f_{sed}\, )$ where $\Delta z_s$ is unit thickness and $N_r$ is the number of rivers. Assuming rivers at indistinguishable stratigraphic levels were active simultaneously, $N_r \approx 20$. With $\Delta z_s \approx 300$ m (Kite et al., submitted), we obtain $\tau = 0.5\text{-}1$ Myr. At long-term terrestrial floodplain aggradation rates $\Delta z_s \approx 300$ m corresponds to a time interval of $1.5 \times 10^4 - 9 \times 10^5$ yr (Bridge and Leeder 1979). These values are all less than the estimate from embedded-crater density, consistent with intermittency, although the error bars are large (e.g., Buhler et al. 2014, Irwin et al. 2015).

It would be interesting to incorporate fluvial intermittency as a constraint on numerical simulations of catchment response to climate change (Armitage et al. 2013) and to investigate whether lower limits on the duration of Mars river activity can be inferred from the spatial scales at which cut-and-fill cycles are correlated across catchments and across the basin.

## 3.4 A tool to search for abrupt climate change on Mars.

To search for paleohydrologic evidence of stratigraphically abrupt climate change on Mars, we analyze our dataset to see if it contains a sharp transition in river-deposit dimensions that is robust against the errors captured by our bootstrap procedure. To do this, we use the K-S test to find breakpoints for each of an ensemble of bootstrapped datasets. For each bootstrapped dataset, we define $z_s^*$ as the $z_s$ that minimizes the K-S statistic (i.e., the $z_s$ corresponding to the most statistically-significant breakpoint between data above $z_s$ and data below $z_s$). We find $z_s^*$ for every bootstrapped distribution of data in the ensemble. If abrupt climate change had occurred, then $z_s^*$ would vary little between the members of the ensemble – it would be tightly clustered in $z_s$ (Fig. 7b). If instead climate change was only gradual, then we would expect to see a broad distribution of $z_s^*$ (Fig. 7b).

Wide error bars can produce false negatives for abrupt climate change. To demonstrate that our dataset is rich enough to detect an abrupt climate change (had it occurred), we generate a synthetic dataset that has the same error bars and $z_s$ values as the observations, but has $\lambda$ and $w$ set to a uniform value below $z_s = 0$ m and set to a different uniform value above $z_s = 0$ m. We recover a sharp break in both $\lambda$ and $w$ at $z_s \approx 0$ m after passing the synthetic dataset through our algorithm (Fig. 7c), which shows that for this dataset our abrupt-change detection scheme is unbiased and robust to false negatives. Finally, we use the observations (Fig. 7d). (Transect 3 data points were excluded). We find that $z_s^*$ is fairly widely scattered in both $\lambda$ and $w$ (5%-95% range of $z_s^*$ is 35m for $\lambda$ and 90m for $w$). We conclude that our stratigraphic logs do not require stratigraphically abrupt climate change on Mars.

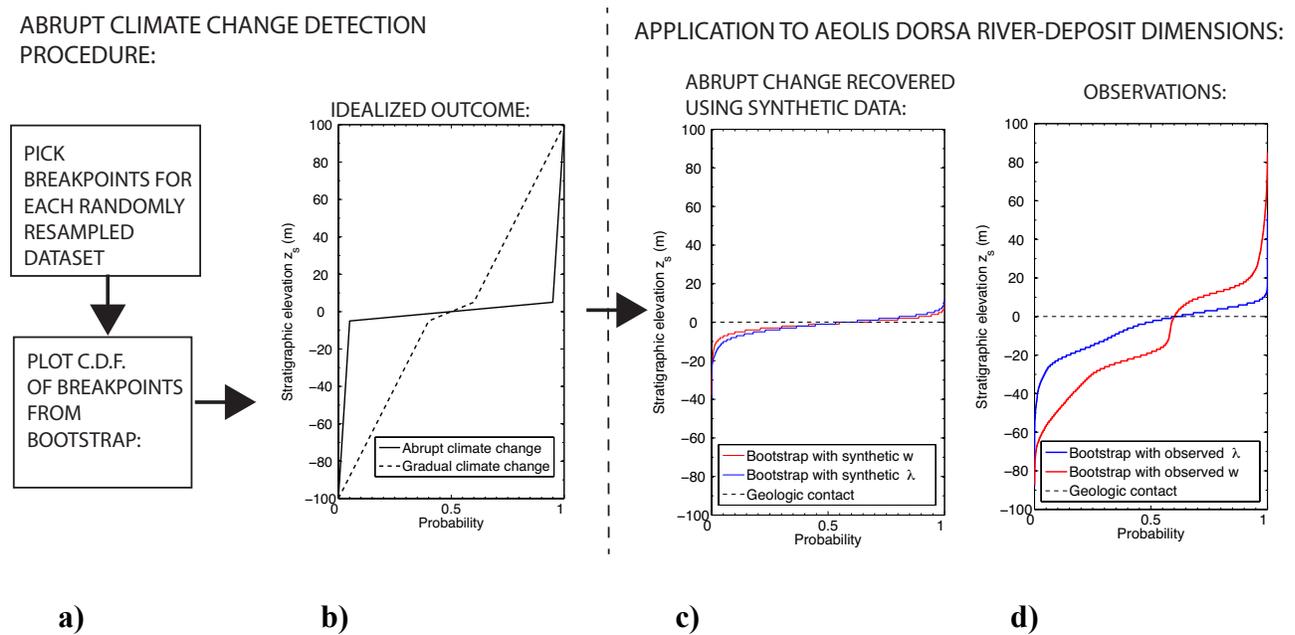

**Figure 7. a)** Abrupt climate change detection concept. **b)** Idealized outcomes. **c)** Application to synthetic dataset with real errors and stratigraphic elevations, and artificially-sharp breakpoints in $\lambda$ and $w$. **d)** Application to Aeolis Dorsa observations.

# 4. Implications for river-forming paleo-climates on Mars.

## 4.1. Mechanisms for paleodischarge variability.

Our preferred explanation for the $\{w, \lambda\}$ changes is a decrease in runoff production (i.e., climate change) (Foreman et al. 2012). The required *relative* change in runoff production for constant drainage area $A_d$ is a factor of ~2.6 (Table 1) ($A_d = L/d_d$, where $L$ is the distance to the drainage divides and $d_d$ is drainage density). An estimate of $A_d$ allows constraints on absolute changes in runoff production. For R-1, a lower bound on $L$ is the distance to the tips of the presently-preserved channel network (~40 km); an upper bound is the distance to the dichotomy scarp (~200 km to the S). The R-1 drainage density is assumed constant at 0.2 km$^{-1}$. For R-2, a lower bound on $L$ is the length of currently-exposed channels (~10 km). The upper limit is again the distance to the dichotomy scarp. The R-2 drainage density is assumed constant at 1 km$^{-1}$. More precise estimates are difficult because erosion has removed the drainage divides of the catchments feeding the rivers preserved in our transects. Near-surface hydraulic conductivity is assumed not to change with time (not unreasonable in an aggrading system). We express the required change in runoff production (Fig. 8) in terms of the energy available for melting snow/ice $E_{melt} = Q \rho c / A$, where $c$ is the latent heat of melting ice (334 kJ/kg) and $\rho$ is the density of water. (None of the data presented in this study argues against rainfall providing the runoff for the Aeolis Dorsa rivers; using $E_{melt}$ is simply a convenient way of expressing the energy budget). The preferred drainage area is ~400 km$^2$, for which peak melt energy (peak runoff production) decreases from 63-83 W/m$^2$ (dividing this value by the latent heat of melting ice and the density of water gives 0.7-0.9 mm/hr) for R-1, to 23-31 W/m$^2$ (~0.3 mm/hr) for R-2. For the relatively large catchments studied here, this should probably be interpreted as a constraint on daily-average runoff production. The absolute magnitude of change scales inversely with $A_d$, and because $A_d$ is poorly constrained we regard the absolute magnitude with some suspicion. However, a reduction of 50 W/m$^2$ in $E_{melt}$ could result from a number of non-exotic processes. Examples include an rise in albedo from 0.2 to 0.3, a decrease in peak eccentricity from 0.14 to 0.1, or a decline in atmospheric pressure from ~100 mbar to ~40 mbar with a corresponding increase in evaporitic cooling (Kite et al. 2013a, Mischna et al. 2013).

Three-fold $Q$ variability does not exclude quasi-periodic orbital forcing. Small variations between successive peaks in orbital forcing can lead to large differences in $Q$. Thresholds amplify input variations (Huybers & Wunsch 2003), and there are multiple thresholds in the chain of processes linking insolation to carving a river. For example, melting initiates at 273 K, overland flow only occurs when melt exceeds infiltration capacity, and sediment movement begins above a critical Shields stress.

Inferred $Q$ variations are completely inconsistent with monotonically decreasing rainfall following an impact, but more sophisticated models of impact-triggered rainfall show seasonal variations (Segura et al. 2013). If an impact triggered a metastable greenhouse (e.g. Urata & Toon 2013, Segura et al. 2013) then season and orbital variability would be expected.

A change in $w$ need not imply a change in $Q$ – if slope $S$ steepens, rivers flow faster, and the same discharge is conveyed by a narrower channel. Field measurements and theory show that $w \propto S^{-3/16}$ (Lee & Julien 2006, Finnegan et al. 2005). Therefore, a 100-fold increase in $S$ between R-1 and R-2 could cause the ~2 fold reduction in $\{w, \lambda\}$. While recognizing the potential complication of post-depositional tilting (Lefort et al. 2012, Lefort et al. 2015), the present-day tilt is ~0.01 to the north for R-2 channels in our DTMs, so supposing $S = 10^{-4}$ for R-1 channels, the 100-fold increase in $S$ between R-2 and R-1 channels can explain $w$ reductions. In order for $S$ to increase within an aggrading sediment package in the absence of changes in discharge or changes in tectonic boundary conditions, either a coarsening of input grain-size or an increase in the rate of non-fluvial sediment input is required. (Flexure under the sediment load would most likely lead to southward tilting, which has the wrong sign to explain the observations.) Although slope increase may have occurred, we do not believe this is sufficient to explain the observed narrowing because a slope of 0.01 over the 40 km N-S extent of R-2 outcrop implies a wedge of sediment thickening to 0.4 km. No such wedge is observed; instead, additional units lacking embedded fluvial deposits are exposed in scarps to the S of the area shown in Fig. 2b (Kite et al., submitted) and we interpret these units to have formerly extended out over the area of our transects.

Another scenario in which $w$ is reduced without a change in $Q$ is if $A_d$ is reduced upsection. In this picture, aeolian sediment roughens the landscape between R-1 and R-2 time (an Earth example is described in Haberlah et al. 2010) and breaks up the smooth alluvial R-1 landscape into small catchments, and these small catchments are not integrated by the R-2 rivers. The R-2 channels have a network pattern that is hard to reconcile with the aeolian-roughening hypothesis. Drainage along linear interdunes would leave a linear channel pattern, but the R-2 networks lack this pattern. We also cannot rule out reduction in $A_d$ due to drainage capture by catchments to the south. These $A_d$–change scenarios assume paleo-drainage to the N in R-1/R-2, which is our preferred interpretation in part because the regional tilt is to the north. If paleo-drainage was instead directed to the south, then variation in $A_d$ is even less likely because river-long-profile distances in Aeolis Dorsa can exceed >500 km (Williams et al. 2013); such large catchments would evolve slowly (Armitage et al. 2013).

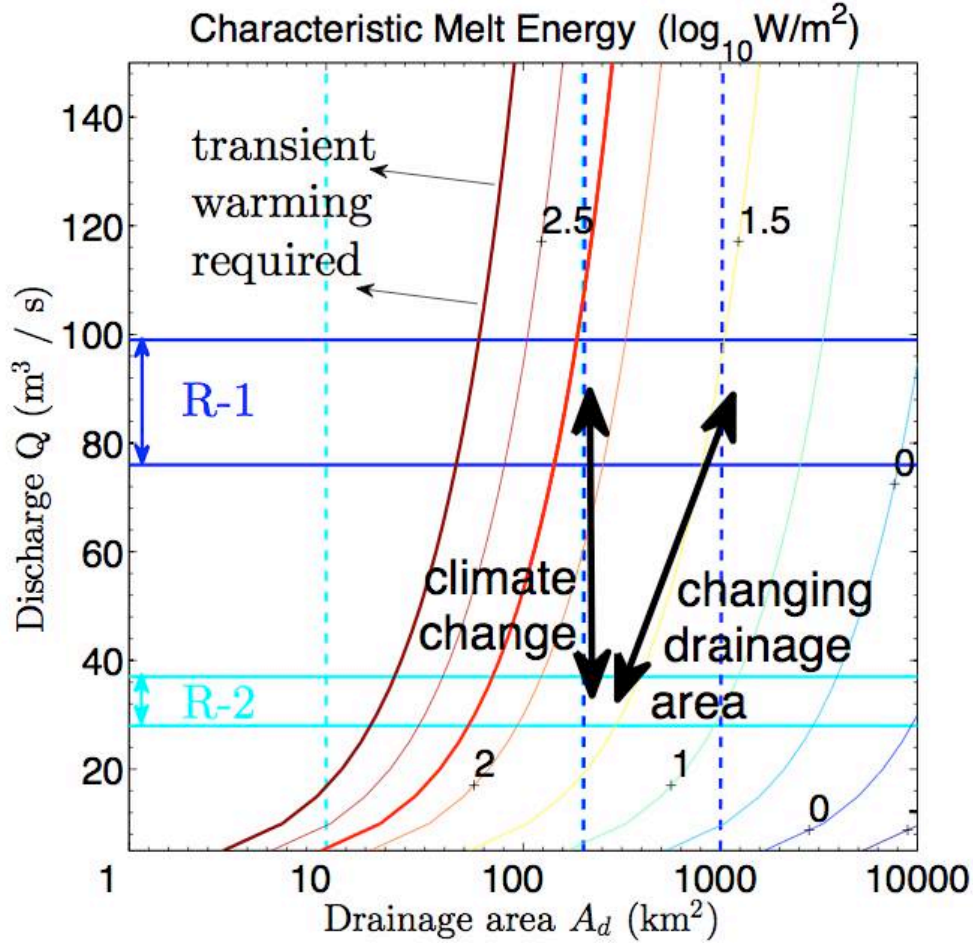

**Figure 8.** The horizontal lines show runoff constraints for R-1 and R-2, and the vertical dashed lines show estimated drainage areas. The thick red curve shows diurnally-averaged equatorial perihelion equinox insolation for an eccentricity of 0.12 and solar luminosity of 80% that of present. The thick brown curve shows noontime equatorial perihelion equinox insolation for an eccentricity of 0.12, solar luminosity 80% that of present. The discharge curves make the unrealistic assumption that melt energy is equal to insolation, and so serves as an upper limit on insolation. There is also the issue of flow routing through the drainage network, tending to reduce flood peaks, and the necessity for discharge to be routed over distances >10 km. The thin colored curves are labelled with $\log_{10}$(melt energy) in units of $\log_{10}(W)/m^2$. The black arrows show pathways for climate change (vertical black arrow) and changing drainage area (oblique black arrow).

## 4.2. Global context.

Our Aeolis Dorsa results add strength to an emerging global picture of time-variable, intermittent, occasionally voluminous river discharge driven by wet paleoclimates and extending remarkably late in Mars history (e.g. Weitz et al. 2013, Grant & Wilson 2011, Fassett et al. 2010, Williams & Weitz 2014). Intermittency is required to explain closed crater-lake basins containing deltas and fed by channels with measured widths because continuous flow would overspill the closed basin, incise an exit channel, and breach the crater rim (Matsubara et al. 2011, Irwin et al. 2007, Lewis & Aharonson 2006, Buhler et al. 2014, Mangold et al. 2012). Fluvial intermittency is also required to account for indurated aeolian bedforms interbedded with river/lake deposits (Milliken et al. 2014, Willliams & Weitz 2014). Large alluvial fans prograding into crater floors lacking deep paleolakes also set limits on discharge magnitude and duration (e.g., Morgan et al. 2014). Our 0.5-1 Myr estimated duration of wet conditions is consistent with estimates elsewhere on Mars (using erosional timescales, e.g. Hoke et al. (2011), or sedimentary deposits, e.g. Armitage et al. (2011)). However, because of erosion and burial, these estimates could correspond to totally distinct episodes. Mineralogic constraints complement geomorphic constraints. Mineralogy suggests pervasive, generally isochemical alteration at low water/rock ratio with preferential dissolution of olivine (e.g. Olsen & Rimstidt 2007, Hurowitz & McLennan 2007, Berger et al. 2009, Elwood Madden et al. 2009, Bandfield et al. 2011, Ehlmann et al. 2011, McLennan et al 2014). In particular, orbiter detections of aqueous minerals associated with well-preserved fluvial features are limited to opal and detrital clays (e.g., Milliken & Bish 2010).

    An open question is whether the probability density function (pdf) of post-Noachian Mars surface habitability was bimodal or unimodal. In the unimodal hypothesis, post-Noachian rivers record orbital-forcing optima of a long-term climate evolution driven by unidirectional atmospheric loss and increasing solar luminosity and perturbed by stochastic orbital variability. The unimodal hypothesis is attractively simple, but sets a challenge for the modeling community:

can a "wet tail" climate-evolution scenario that generates *peak* runoff high enough to account for rivers like those in Aeolis Dorsa avoid breaching mineralogical upper limits on *total* liquid water availability (e.g. Tosca & Knoll 2009)? If not, then late-stage rivers would require a bimodal climate evolution pdf – in which post-Noachian rivers record a discrete class of (metastable?) more-habitable climates.

# 5. Habitability, taphonomy, and organic matter preservation.

The search for life on Mars has narrowed to a search for ancient biogenic organic matter (Farmer & DesMarais 1999, Grotzinger 2014). In order for ancient complex biogenic organic matter to be present in the near-surface today, the paleo-environment must have been habitable, with high organic-matter preservation potential; and the rocks must have been rapidly exhumed to minimize radiolysis. In Aeolis Dorsa, meander cutoffs define oxbow lakes, which are favored for the preservation of fine-grained sediments that tend to bind organic matter (Constantine et al. 2010, Ehlmann et al. 2008). River floodplains have better organic-matter preservation potential than aeolian deposits, alluvial-fan deposits, or fluvial-channel deposits, although distal lacustrine sites are better still (Summons et al. 2011, Williams & Weitz 2014). Aeolis Dorsa river discharges and flow durations suggest a habitable environment, the relatively high deposition rates favor preservation against early degradation, and the relatively high erosion rate protects against radiolysis during exhumation. In particular, a 3000 km$^2$ area around Transect 1 lacks craters larger than $D > 1$ km, corresponding to a nominal erosion rate of 0.2-0.4 μm/yr (using the chronology of Werner & Tanaka, 2011). This is sufficiently high to minimize radiolysis of complex organic matter at the depths sampled by Mars 2020 (several cm) and the ExoMars rover (≤2 m) (Pavlov et al. 2012, Farley et al. 2014). Therefore the river deposits we document are a promising alternative for paleobiological exploration in the event that suitable lacustrine deposits can not be identified or sampled in situ.

# 6. Conclusions.

We analyze past river processes at Aeolis Dorsa, Mars, by collapsing measurements of river-deposit dimensions onto stratigraphic logs[4]. Our stratigraphic logs show a (1.5-2)× reduction in river-deposit dimensions between two river deposits. This is consistent with a 2.6× reduction in peak discharge across the contact, using size-discharge scalings modified for Mars gravity. Marker-bed stratigraphy suggests additional variations in peak discharge at stratigraphic scales below the resolution of the logs. The total time interval for these changes probably exceeded 0.5 Myr. Similar to the Grand Staircase, Utah, USA, at Aeolis Dorsa we see multiple layers of ancient river deposits exposed by modern erosion. The requirement for intermittency at multiple timescales (as shown by river-deposit dimensions, regional unconformities, and marker beds) is a stringent constraint on quantitative models linking fluvial sedimentology to late-stage climate evolution.

# Acknowledgements.

This paper grew from discussions in the Caltech Mars Fluvial Geomorphology Reading Group, and we thank all the participants, especially Mike Lamb and Roman DiBiase. To the extent we understand Aeolis Dorsa, it is thanks to conversations with the following scientists, who we have relied heavily on in thinking through this problem: Alexandra LeFort, Laura Kerber, Caleb Fassett, Noah Finnegan, Lynn Carter, Nicolas Mangold, Sanjeev Gupta, Matt Balme, Sam Harrison, Maarten Kleinhans, Leif Karlstrom, David Mohrig, Benjamin Cardenas, Jim Zimbelman, Steve Scheidt, Michael Manga, Bill Dietrich, Ross Irwin, Jeff Moore, Christian Braudrick, Devon Burr, Robert Jacobsen, Noah Finnegan, Jonathan Stock, Gary Kocurek, Richard Heermance, Paul E. Olsen, Or Bialik, Brian Hynek. We thank Robert Jacobsen for comments on a draft. E.S.K. thanks Ross Beyer, Sarah Mattson, Annie Howington-Kraus, and Cynthia Phillips for help with generating the PSP_006683_1740/PSP_010322_1740 DTM. E.S.K. was supported by a Princeton University fellowship and by NASA grant NNX11AF51G. We thank the HiRISE team for maintaining the HiWish program, which supplied multiple images essential for this study.

---

[4] This technique may be particularly useful for extracting records from High-Resolution Stereo Camera (HRSC)/Mars Express data and from Color and Stereo Surface Imaging System (CaSSIS)/ExoMars Trace Gas Orbiter data.

# References.

Leeder, M. R., Harris, T. and Kirkby, M. J. (1998), Sediment supply and climate change: implications for basin stratigraphy. Basin Research, 10: 7–18.

Lefort, A., et al. (2012). Inverted fluvial features in the Aeolis-Zephyria Plana, western Medusae Fossae Formation, Mars: Evidence for post-formation modification. J. Geophys. Res. (Planets) 117, 3007.

Lefort, A. et al. (2015), Channel slope reversal near the Martian dichotomy boundary: Testing tectonic hypotheses, doi:10.1016/j.geomorph.2014.09.028.

Lewis, K.W., & Aharonson, O. (2006), Stratigraphic analysis of the distributary fan in Eberswalde crater using stereo imagery, J. Geophys. Res. 111, E6, E06001.

Macklin, M.G., J. Lewin, and J. C. Woodward, (2012) The fluvial record of climate change, Phil. Trans. R. Soc. A May 13, 2012 370 1966 2143-2172; doi:10.1098/rsta.2011.0608 1471-2962

Malin, M.C., et al. (2010), An overview of the 1985-2006 Mars Orbiter Camera science investigation, International Journal of Mars Science and Exploration (Mars Journal), 4, 1-60.

Mangold, N., et al. (2004), Evidence for Precipitation on Mars from Dendritic Valleys in the Valles Marineris Area, Science, 305, 78-81.

Mangold, N., et al. (2012), The origin and timing of fluvial activity at Eberswalde crater, Mars, Icarus, 220, 530-551.

Matsubara, Y.; Howard, A.D.; Drummond, S.A. (2011), Hydrology of early Mars: Lake basins, J. Geophys. Res., 116, Issue E4, CiteID E04001.

Matsubara., Y. et al. (submitted), River meandering without vegetation on Earth and Mars: A comparative study Aeolis Dorsa meanders, Mars and possible terrestrial analogs of the Usuktuk River, AK, and the Quinn River, NV.

McLennan, S.M., et al. (2014), Elemental Geochemistry of Sedimentary Rocks at Yellowknife Bay, Gale Crater, Mars, Science, 343 (6169), 1244734.

McNamara, J.P., & Kane, D.L. (2009), The impact of a shrinking cryosphere on the form of arctic alluvial channels, Hydrological Processes, 23, 159-168.

Metz, J.M., et al. (2009), Sulfate-Rich Eolian and Wet Interdune Deposits, Erebus Crater, Meridiani Planum, J. Sediment. Res., 79, 247-264.

Milliken, Ralph E.; Bish, David L. (2010), Sources and sinks of clay minerals on Mars, Philosophical Magazine, vol. 90, issue 17, pp. 2293-2308.

Milliken, R. E.; Ewing, R. C.; Fischer, W. W.; Hurowitz, J. (2014), Wind-blown sandstones cemented by sulfate and clay minerals in Gale Crater, Mars, Geophys. Res. Lett., 41, 1149-1154.
38

Mischna, M. et al. (2013), Effects of obliquity and water vapor/trace gas greenhouses in the early martian climate, J. Geophys. Res.: Planets, 118, 560-576

Morgan, A.M., et al. (2014), Sedimentology and climatic environment of alluvial fans in the martian Saheki crater and a comparison with terrestrial fans in the Atacama Desert, Icarus, 229, 131-156.

Olsen, A., Rimstidt, J. (2007), Using a mineral lifetime diagram to evaluate the persistence of olivine on Mars. American Mineralogist 92 (4), 598–602.

Palucis, M., et al. (2014), The origin and evolution of the Peace Vallis fan system that drains to the Curiosity landing area, Gale Crater, Mars, J. Geophys. Res. 119, 705-728.

Parker, G., et al. (2007), Physical basis for quasi-universal relations describing bankfull hydraulic geometry of single-thread gravel bed rivers, J. Geophys. Res.- Earth Surface, 112, CiteID F04005.

Pavlov, A. A.; Vasilyev, G.; Ostryakov, V. M.; Pavlov, A. K.; Mahaffy, P. (2012), Degradation of the organic molecules in the shallow subsurface of Mars due to irradiation by cosmic rays, Geophys. Res. Lett., 39, CiteID L13202.

Ramirez, R. et al., (2014), Warming early Mars with $CO_2$ and $H_2$, Nat. Geosci., 7, 59-63.

Reijenstein, H.M., et al., (2011), Seismic geomorphology and high-resolution seismic stratigraphy of inner-shelf fluvial, estuarine, deltaic, and marine sequences, Gulf of Thailand, AAPG Bulletin, 95, 1959-1990.

Robbins, S.J.; et al. (2014), The variability of crater identification among expert and community crater analysts, Icarus, 234, p. 109-131.

Sadler, P.M., & D.J. Jerolmack (2014), Scaling laws for aggradation, denudation and progradation rates: the case for time-scale invariance at sediment sources and sinks, *in* Smith, D. G., et al. (eds) Strata and Time: Probing the Gaps in Our Understanding. Geological Society, London, Special Publications, 404, http://dx.doi.org/10.1144/SP404.7.

Schmitz, B., and Pujalte, V. (2007), Abrupt increase in seasonal extreme precipitation at the Paleocene-Eocene boundary, Geology, March, 2007, v. 35, p. 215-218, doi:10.1130/G23261A.1

Segura, T.L., Zahnle, K., Toon, O.B., & McKay, C.P. (2013), The effects of impacts on the climates of terrestrial planets, p. 417-438 in Mackwell, S.J., et al., eds., Comparative Climatology of Terrestrial Planets, U. Arizona Press.

Smith, Matthew R.; Gillespie, Alan R.; Montgomery, David R. (2008), Effect of obliteration on crater-count chronologies for Martian surfaces, Geophys. Res. Lett., 35, Issue 10, CiteID L1020.

# Supplementary Materials.

# A. Supplementary Methods.

## A1. Stratigraphic-elevation ($z_s$) assignments.

Topographic elevations $z$ for Transects 2 and 3 were taken from the HiRISE DTM (Table S1). For Transect 1, we assigned $z$ from a CTX DTM (Table S1), except for a small strip falling off the W edge of the CTX DTM (Figure S1a). To obtain $z$ for this strip, we regressed HiRISE DTM $z$ on CTX DTM $z$ for the overlapping part of the DTMs, and applied the linear trendline to convert HiRISE DTM $z$ to "equivalent" CTX DTM $z$.

The geologic contact between R-1 and R-2 ($z_s = 0$ m) is defined by a change in erosional expression and by the disappearance of large meander belts. We picked lines where the meander belts disappear beneath R-2 cover. Each of the meander belts has an approximately horizontal top and is much smaller in width than the DTM width, so we used ArcGIS "polyline to point" with "inside" checked to interpolate each line to a single point with a single $z$ value. We then interpolated the $z_s = 0$ m surface between these measurements using a planar surface, a quadratic polynomial surface, an inverse-distance-weighted (IDW) surface (using the 12 nearest points), and using ordinary kriging with a spherical semivariogram. Our conclusions are not sensitive to the choice of interpolation method, but the results do differ in detail. Which interpolation is best? Quadratic polynomial interpolation is the only method that scores well on both of the following measures:- (1) Does the interpolation assign width measurements that have been tagged with a geologic unit to the correct unit ($z_s < 0$ m for R-1, and $z_s > 0$ m for R-2)? (2) Is the marker bed (§3.2) at a constant stratigraphic elevation? On the first test, both quadratic interpolation and IDW score well (>90% of points assigned to correct unit) while planar interpolation and kriging score less well (<90% of points assigned to correct unit). The marker bed is only exposed in Transect 1. For the marker bed, the stratigraphic elevation traced along the IDW and kriged surfaces varies by >150 m, suggesting poor fit. The stratigraphic elevation of the marker bed varies by <50m when referenced to the the planar and quadratic interpolated surfaces. Therefore we use quadratic interpolation because only the quadratic interpolation does well for both of our tests.[5]

| Transect | Stereopair | DTM resolution (m/pixel) |
|---|---|---|
| 1 | ESP_017548_1740/ESP_019104_1740 | 2 |
| | ESP_019038_1740/ESP_019882_1740 | 1 |
| | PSP_006683_1740/PSP_010322_1740 | 1 |
| 2 | PSP_007474_1745/ESP_024497_1745 | 2.5 |
| 3 | PSP_002002_1735/PSP_002279_1735 | 2 |
| 1 (only used for contact interpolation) | B20_017548_1739_XI_06S206W/ G02_019104_1740_XI_06S206W | 30 |

**Table S1.** The HiRISE and CTX DTMs used in this study. All DTMs were produced by A.S.L., except for PSP_006683_1740/PSP_010322_1740 which was produced by E.S.K.

For Transect 3, the R-1/R-2 contact is not exposed, so we extrapolated the contact from regional mapping (Kite et al., submitted). The following approach assumes that the topographic offset between Transect 3 and R-1 corresponds to a stratigraphic offset – for discussion of alternative hypotheses that are consistent with the data, see the main text and Kite et al. (submitted). Over the whole region shown in Fig. 2b, we extracted all MOLA Precision Experimental Data Record (PEDR) points whose center coordinates fell within 150 m of the hand-picked contact (except where the contact was flagged as "inferred"). We chose 150 m because this is ~1/2 the along-track distance between MOLA PEDR laser shots. For these PEDR points, we fitted a planar surface (RMS error 50 m) and a quadratic surface (RMS error 40 m) to the MOLA elevations and extrapolated both surfaces to three locations within Transect 3, finding extrapolated contact-surface elevations of (-2113 ± 35) m (planar) and (-2012 ± 42) m (quadratic). We also picked the 10 nearest MOLA PEDR points to Transect 3, finding elevation (-2006 ± 86) m. Equally weighting the three extrapolations (quadratic global, planar global, and same-as-nearest-neighbour) we find (-2044 ± 67) m as the base elevation for Transect 3. Errors in *z* (intrinsic DTM errors) are small compared to interpolation errors (errors in $z_s$).

---

[5] For Transect 1, the assumption of isotropy underlying our kriging algorithm is troubling because of preferred wrinkle ridge orientations (obvious in MOLA; see also Lefort et al. 2012). This is an additional strike against using kriging.

Interpreting a $z_s$ series as a relative-time series is only valid if the river deposits are eroding out of the rock (rather than being late-stage unconformable cut-and-fill imprinted on near-modern topography, as at Gale crater; Milliken et al. 2014). Kite et al. (submitted) show that the Aeolis Dorsa river deposits are eroding out of the rock. A second requirement is that the amplitude of incised valleys (cut-and-fill cycles) must be smaller than the scale of data interpretation. Very-large scale incised valleys can be ruled out in southern Aeolis Dorsa (Kite et al., submitted), but it is possible that the R-1/R-2 contact is somewhat time-transgressive.

A common pattern for river-deposit erosional expression in S Aeolis Dorsa is that a plug of sediment is exposed as an inverted channel at higher elevations, which can be traced downslope into double ridges, and then traced further downslope into negative relief (widening channel) preservation. Similar forms of preservation (double ridge, negative relief) are observed at Aeolis Serpens, slightly north of our study area (Williams et al. 2013). This can be interpreted in terms of wind erosion first exhuming, and then progressively obliterating, a channel-filling sediment plug (Farley et al. 2014). This supports our inference that the channel deposits are embedded in stratigraphy.

The outcrop-scale observation that at least some meandering channel systems were aggradational (Fig. 1, Fig. 6) gives confidence that the observed inverted channels are channel deposits. Fluvial sedimentary bedforms would confirm this interpretation, but were not observed in the orbital imagery of these transects. It is possible that some streams were generally incising but interrupted by aeolian deposition that fossilized the channels (such that some channel-filling sediment plugs are aeolian materials and not fluvially transported).

We do not know if the transition seen in Fig. 4 spans the main great drying of Mars or is a brief window into a complicated climate evolution. Evidence for return to wetter conditions high in the stratigraphy favors the latter. Wind or river erosion may have obliterated the sedimentary record of additional, unseen river episodes. There may be an unconformity between R-1 and R-2 (Kite et al., submitted). Correlation between late-stage river events is in its infancy (Ehlmann et al. 2011).

# A2. Paleohydrologic-proxy measurements and locations.

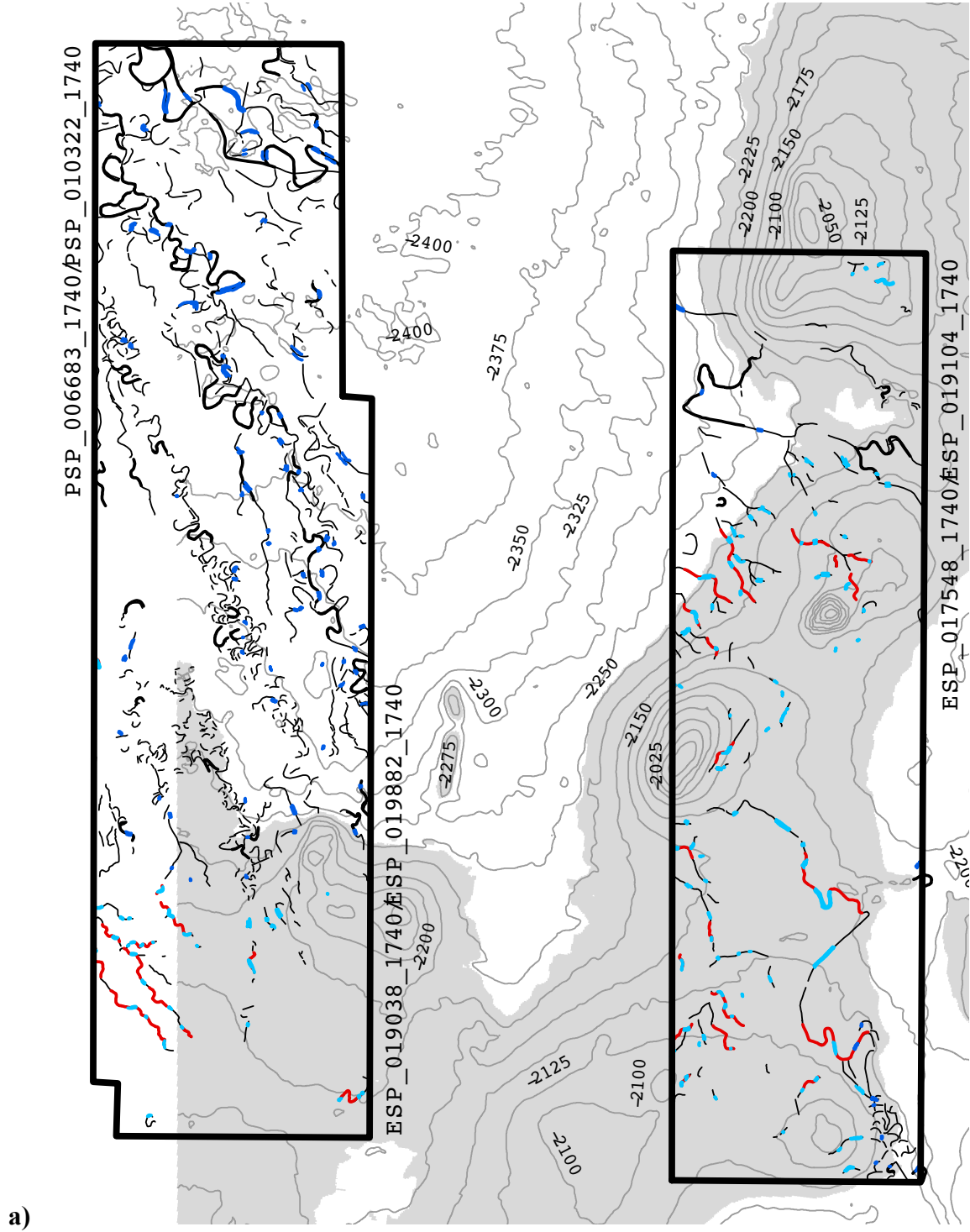

a)

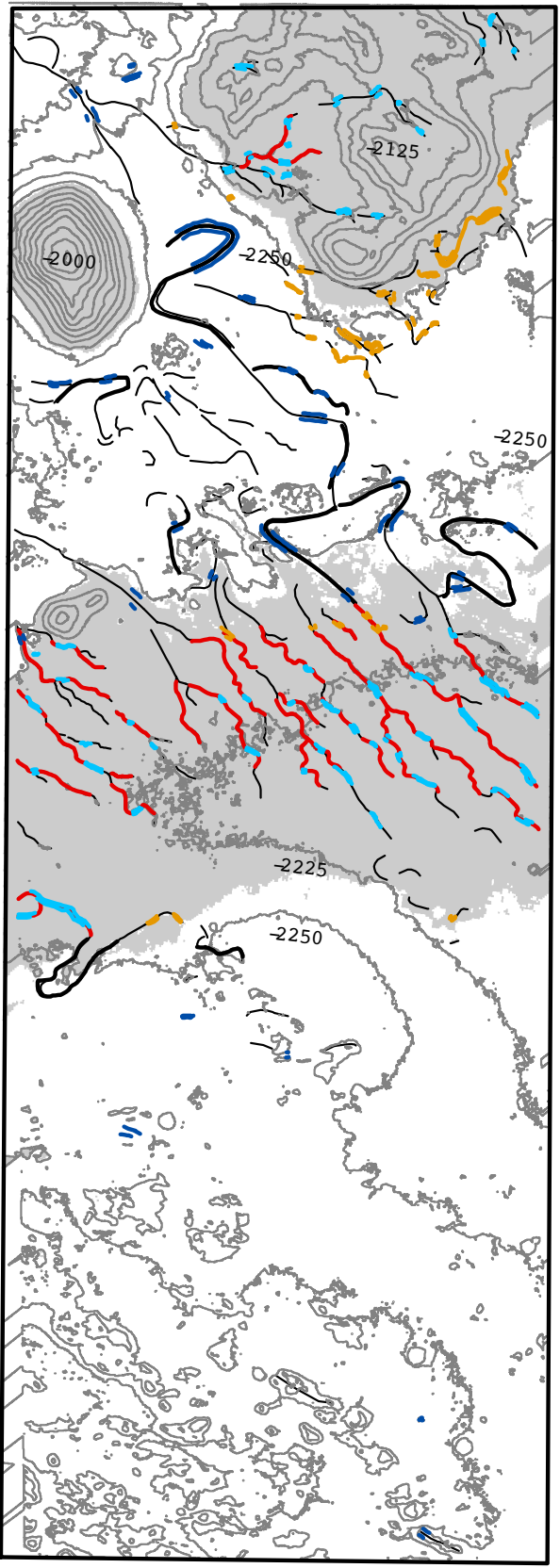

b)

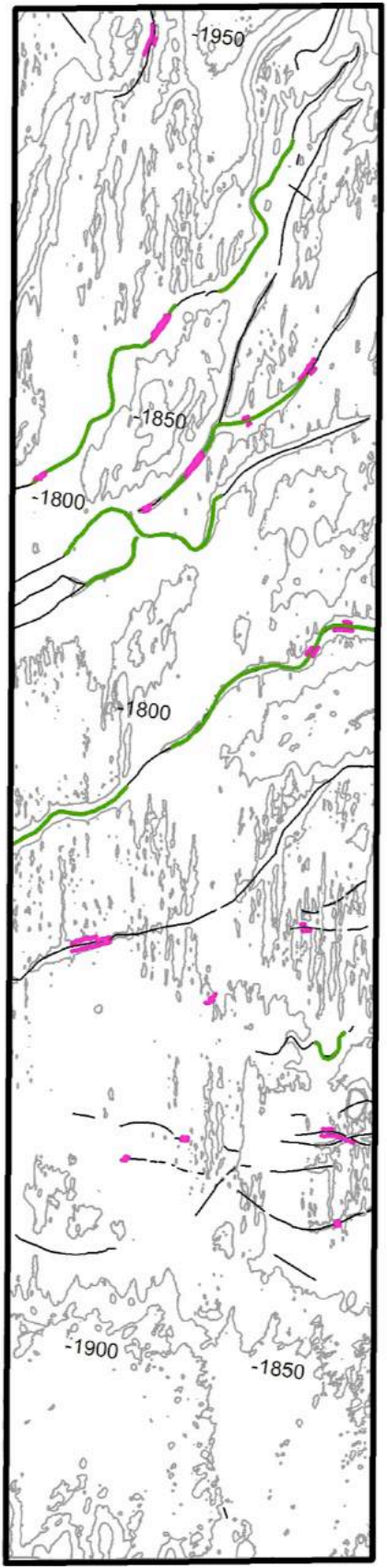

c)

**Figure S1.** Transect-by-transect locator maps showing all measurements. **a)** Transect 1 (gray contours at 25m intervals from CTX DTM; gray shaded area shows area above the interpolated R-1/R-2 contact); **b)** Transect 2 (gray contours at 50m intervals from HiRISE DTM; gray shaded area shows area above the interpolated R-1/R-2 contact); **c)** Transect 3 (gray contours at 25m intervals from HiRISE DTM). For all transects, thin lines correspond to channel threads. Thick colored lines show measured channel segments (centerlines for $\lambda$ and bank-pairs for $w$), as follows:- Dark blue: R-1, $w$. Light blue: R-2, $w$. Pink: Transect 3, $w$. Black: R-1, $\lambda$. Red: R-1, $\lambda$. Green: Transect 3, $\lambda$. Orange shows $\lambda$ and $w$ measurements very close to the R-1/R-2 contact that were not assigned to a unit.

We systematically surveyed our DTMs for paleohydraulic proxies. Survey boxes were defined such that each pixel on the screen corresponded to 1.5 pixels on the orthorectified image. Each survey box was inspected in turn. Channels are preserved in both negative relief (as valleys) and as inverted channels. For the three independent channel-centerline picks, the view was repicked 3 times and rotated by 90° between each pick. The stretches on both the DTM (rainbow color scale) and the orthophoto (grayscale) were set to 2 standard deviations. Complete blinding is not possible because the human operator can generally tell (from the erosional expression of the river deposits) what part of the stratigraphic column they are working in. However, because almost all channel picks were done at a zoom level where the margins of the 'boxes' were not visible, the human operator is usually blind to the absolute scale of the features being picked.

In order to measure $w$ and $\lambda$ as objectively as possible, we extract $w$ and $\lambda$ semi-automatically. The remaining subjective step is the initial selection of channel-centerlines and bank-pairs for tracing, and the assignment of a quality score to each (low-quality "candidate" data are excluded from the fit). The HiRISE DTMs were valuable for this step because they allow us to identify places where channels appeared to be continuous in plan-view HiRISE images, but had large stratigraphic offsets. We used terrestrial compilations to determine the range of acceptable along-channel stratigraphic offsets (Gibling 2006).

The following attributes were assigned for every bank-pair:

| Table S2. Attributes recorded for bank pairs (subsequently processed to extract $w$). | | |
|---|---|---|
| Attribute | Description | Values (*summed for hybrid/combined/intermediate values*) |
| **DoubRidges** | Are double ridges present? | 1 – Yes, double ridges (sensu Williams et al. 2013) are present for this segment of channel (not necessarily the section of bank measured). <br> -1 – No, double ridges are absent. |
| **PQuality** | Preservation quality | 1 – Gold standard. Unusually unambiguous (e.g., paired ridges on <u>banks measured</u> and an inverted channel (topo & ortho). <br> 2 – Good (can be topo <u>or</u> ortho) <br> 4 – Probable. <br> 8 – Candidate. Do not use, but retain for later review. |
| **PStyle** | Preservation style | 1 – Valley (negative relief). <br> 2 – Part of clear meander / sinuous shape (visible lateral-accretion deposits not necessary) <br> 4 – Negative-relief valley containing positive relief deposit. <br> 8 – Inverted channel. <br> 16 – Double ridges only. |
| **DefnBank** | How is measured bank defined? | 1 – Break-in-slope picked using illumination on orthorectified image. <br> 2 – DTM-picked break-in-slope. <br> 4 – Ortho inner channel (e.g. PStyle = 3) or ortho bank indicators (e.g. double ridges or abrupt end of lateral-accretion deposits). <br> 8 – Multiple DTM picks on break-in-slope <u>or</u> trace on curvature raster (from DTM). <br> 16 – Same elevation as break-in-slope on paired bank. |
| **IsSinuous** | Is bank-pair measured on a sinuous channel? | 1 – Definitely has channel-like sinuosity. <br> 0 – Unclear or ambiguous evidence for channel like sinuosity. <br> -1 – No meaningful evidence for channel-like sinuosity. |
| **TieNumber** | | Unique integer joining non-independent bank-pairs (which are resampled together in the bootstrap). |
| **UnitAssign** | Geologic unit hosting measured bank-pairs | 1 – R-1 <br> 2 – R-2 <br> -9 – Uncertain. |
| **BankCode** | | Unique integer for pairs of nearby banks in otherwise confusing terrain. |

The following attributes were recorded for channel centerlines:

**Table S3.** Attributes recorded for channel centerlines (subsequently processed for $\lambda$ measurements).

| Attribute | Description | Values (*summed for hybrid/combined/intermediate values*) |
|---|---|---|
| **PQuality** | Preservation quality | **1** – Gold standard. Unusually unambiguous (e.g., paired ridges on <u>banks measured</u> <u>and</u> an inverted channel (topo & ortho). <br> **2** – Good (can be topo <u>or</u> ortho) <br> **4** – Probable. <br> **8** – Candidate. Do not use, but retain for later review. |
| **PStyle** | Preservation style | **1** – Valley (negative relief). <br> **2** – Part of clear meander / sinuous shape (visible lateral-accretion deposits not necessary) <br> **4** – Negative-relief valley containing positive relief deposit. <br> **8** – Inverted channel. <br> **16** – Double ridges only. |
| **TrAmbig** | Trace ambiguity | In areas of complex preservation: <br> **1** – No significant ambiguity in channel trace. <br> **2** – Single thread, but significant <u>stratigraphic</u> ambiguity. <br> **4** – Single channel, but confusion due to switch between preservation styles (or similar). <br> **8** – Ambiguity between multiple channels or meanders on a single "level" of preservation. <br> **16** – Ambiguity between multiple levels (distinct threads per channel). |
| **TrQual** | Trace quality | Level of "worry" associated with effect of TrAmbig on integrity of meander $\lambda$ measurements. (Analogous to PQuality in Table S2). <br> **1** – $\lambda$ <u>very</u> unlikely significantly affected. <br> **2** – $\lambda$ possibly corrupted at <30% level. <br> **4** – $\lambda$ possibly corrupted at >30% level. <br> **8** – $\lambda$ corruption is likely (do not use, but retain for later review). |
| **UnitAssign** | Geologic unit hosting measured bank-pairs | **1** – R-1. <br> **2** – R-2. <br> **-9** – Uncertain. |
| **SureMeander** | | **1** – Measurements definitely pertain to one or more meanders. <br> **-1** – Measurements do not definitely pertain to one or more meanders. |
| **TieNumber** | | Unique integer joining non-independent channel centerlines (which are resampled together in the bootstrap). |

We also picked polygons outlining areas of channel deposits and evidence for channel migration during aggradation.

# A3. Data reduction - extraction of wavelength (λ) and width (w).

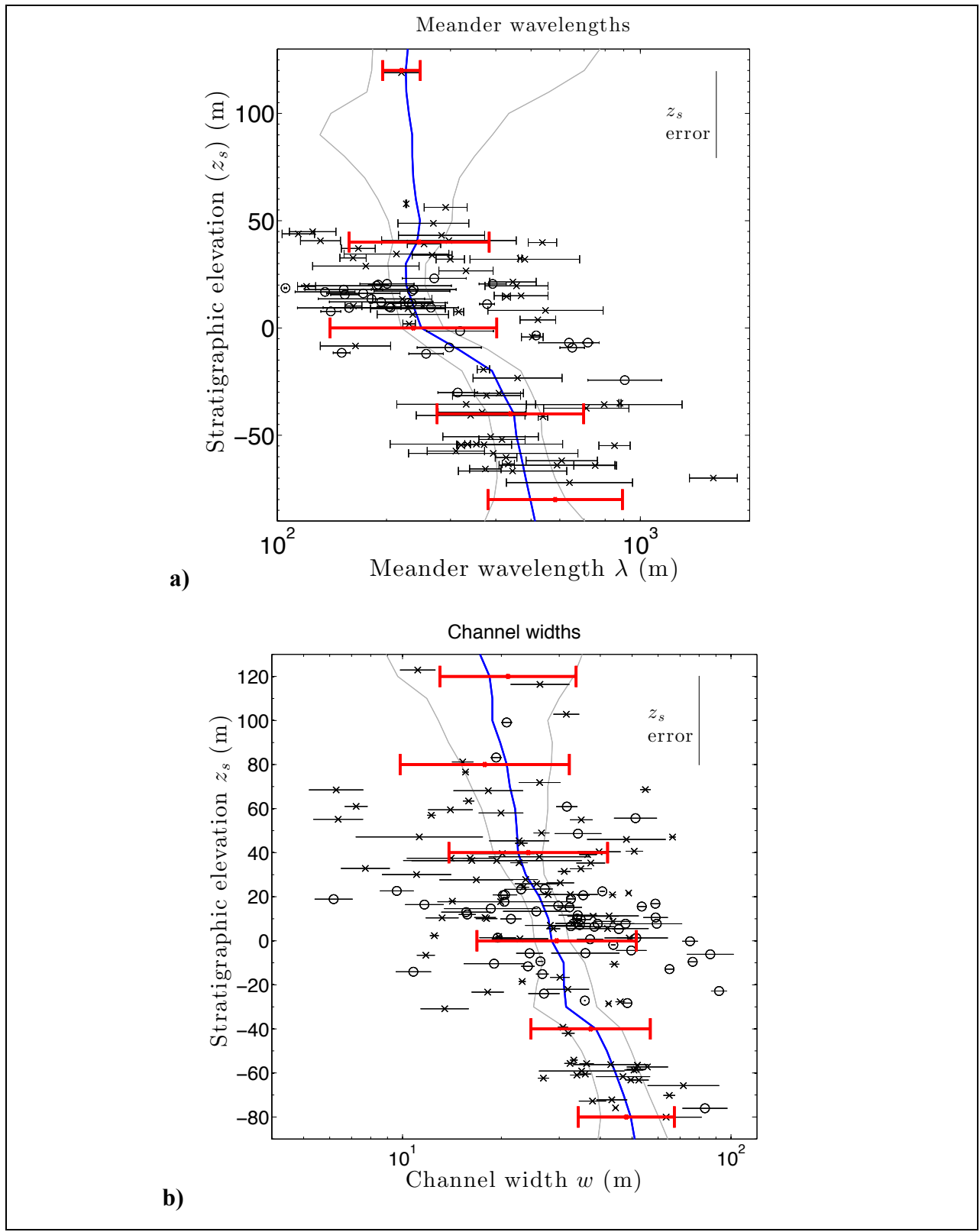

a)

b)

**Figure S2:** Zoom in to the data figures shown in Figure 4, emphasizing the stratigraphy spanning the R-1/R-2 contact ($z_s$ = 0 m). Symbols correspond to transects with crosses × for Transect 1, and circles ○ for Transect 2.

Our data reduction is carried using MATLAB scripts. Paleohydrologic-proxy line picks (bank pairs and channel centerlines) are converted to $w$ and $\lambda$ using `parseMeanderWavelengthMeasurementsMarsRivers` and `parseWidthMeasurementsMarsRivers`. Similarly, area picks are converted to fractional areas versus stratigraphic elevation using `parseAuxiliaryDataMarsRivers`. We do not correct for the fractal dimension of the features being measured, because most of the river channels are parts of networks that have a lateral extent much greater than that of the features being measured (and usually larger than the length of our transects).

Wavelength ($\lambda$) extraction (`parseMeanderWavelengthMeasurementsMarsRivers.m`) :-

This script is heavily influenced by Howard & Hemberger (1991). Channel-centerline picks are ingested from ArcGIS-exported shapefiles. Candidate (low quality) data are then deleted, and points are interpolated uniformly along the channel-centerline polyline traces. The distance between interpolated points is 10 m. The polylines are then converted to coordinates of $\{ s, \theta \}$, where $s$ is distance along the channel and $\theta$ is along-channel direction. Curvature $\partial^2\theta / \partial^2 s$ is then calculated for each interpolated point on the channel, and is smoothed using a 10-point moving baseline. Every point where the curvature changes sign is tagged as a candidate inflection point. Trial half-meanders are defined as the segment of channel between these candidate inflection points. The sinuosity $\xi$ for each trial half-meander is defined as the ratio of along-channel distance between inflection points to the straight-line distance between inflection points. Only half-meanders with $\xi \geq 1.1$ are used. If any candidate inflection point is the end-points for two half-meanders with $\xi < 1.1$, then that candidate inflection point is removed and the trial half-meanders are re-calculated.

A real half-meander should be identifiable in repeated picks of the same channel. Therefore, we compare the three picks of each meandering channel to seek half-meanders that are reproduced on multiple picks. We define reproducibility as requiring that the median distance between points dotted uniformly along the straight lines defining half-meanders be less than ¼ of the straight-line length of the half-meanders. The error assigned to the replicable half-meander is then the standard deviation of the straight-line lengths of the reproduced half-meanders on each trace.

Finally, some centerline traces contain multiple replicable half-meanders. Because our centerline traces are fairly short, meander wavelengths on the same centerline trace are not independent. (If we artificially altered the wavelength of an upstream-most half-meander on one of our centerline traces, then the flow pattern for the downstream half-meanders would be greatly affected). Therefore, we combine the (log-)mean of half-meander wavelengths measured on the same centerline trace. Error bars are also combined assuming a log-Gaussian distribution of errors.

Width ($w$) extraction (`parseWidthMeasurementsMarsRivers.m`): -

Bank-pair picks are ingested from ArcGIS-exported shapefiles. Points are dotted uniformly along each bank. The distance between dots is 2.5m. For each dot, the closest distance to any point on the polyline of the opposite bank is found. (If the closest point is at the *end* of the bank trace, that width measurement is excluded). The mean of these closest distances for the bank-pair is the channel width, and the standard deviation of these closest distances is the error. Bank-pairs measured on the same channel are aggregated, assuming Gaussian errors.

Auxiliary data extraction (`parseAuxiliaryDataMarsRivers.m`): -

Polygons outlining areas of channel deposits and evidence for channel migration during aggradation are ingested from ArcGIS. Raster grids of DTM topography and grids of interpolated stratigraphic surfaces are also ingested from ArcGIS. $z_s$ is set for every point on the grid by subtracting the interpolated stratigraphic surfaces from the topography. Channel polygons are divided into small triangles using Delauney triangulation. The area of each small triangle is calculated. The stratigraphic elevation of the vertices of each small triangle is obtained from the $z_s$ grid. The $z_s$ of each small triangle is assumed to be the mean $z_s$ of the vertices of that triangle. The area of the small triangle is then added to the total area for that stratigraphic-elevation bin. Finally, the total areas for each bin are divided by the exposed outcrop areas (for the entire DTM), found from the raster grids. $z_s$ errors are not taken into account for auxiliary data.

## B. Transect-by-transect stratigraphic logs.

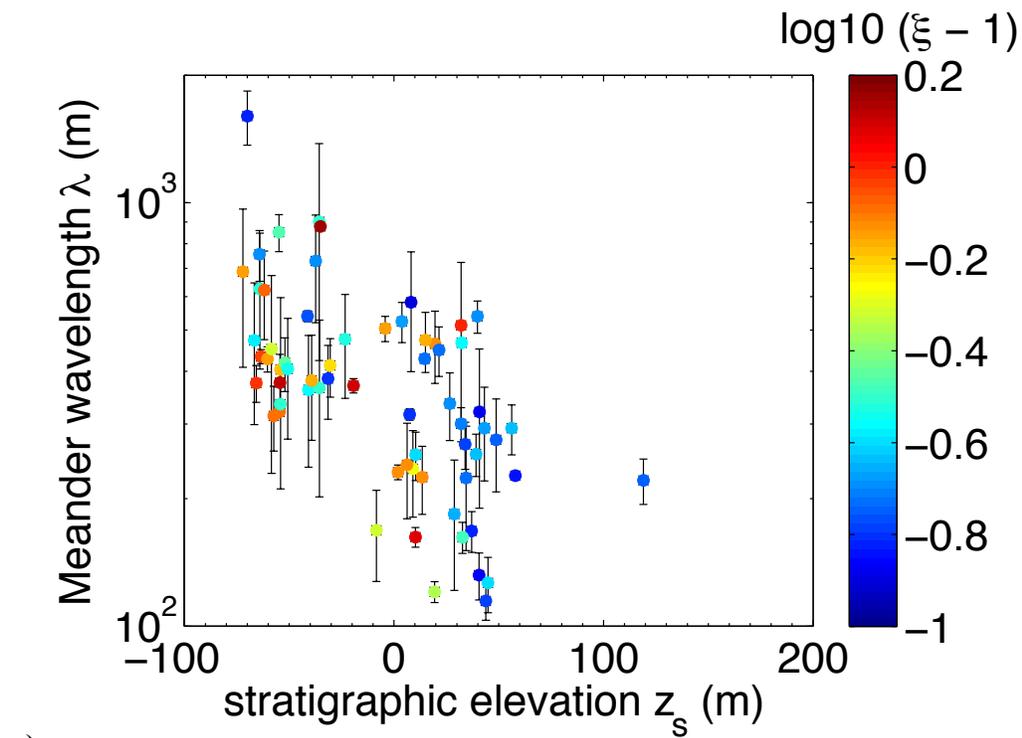

a)

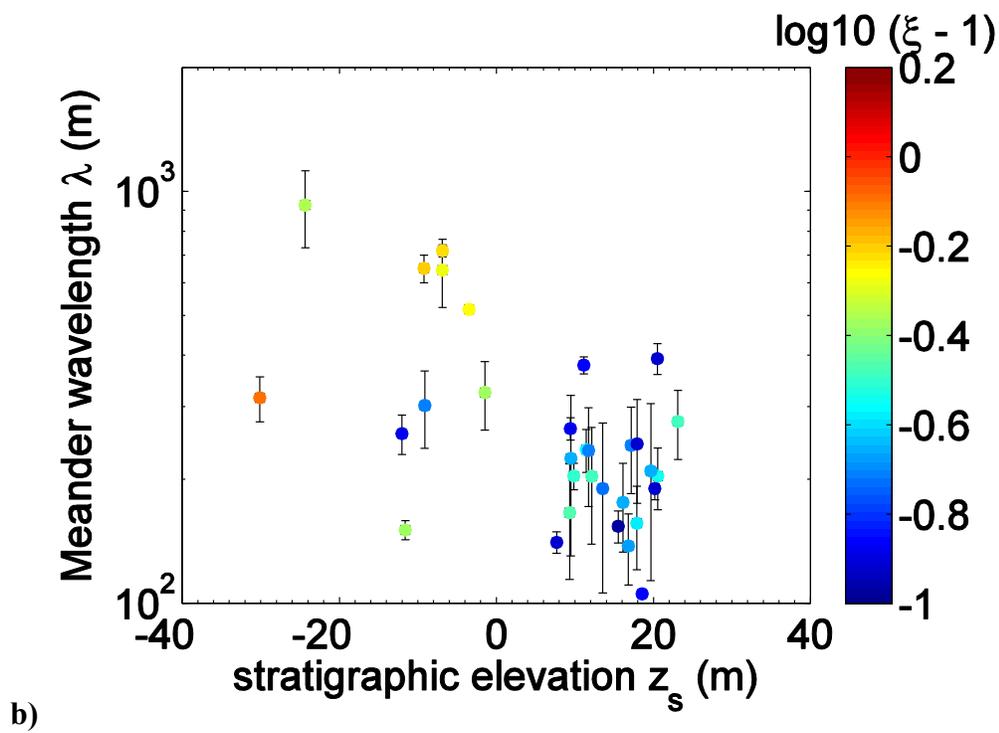

b)

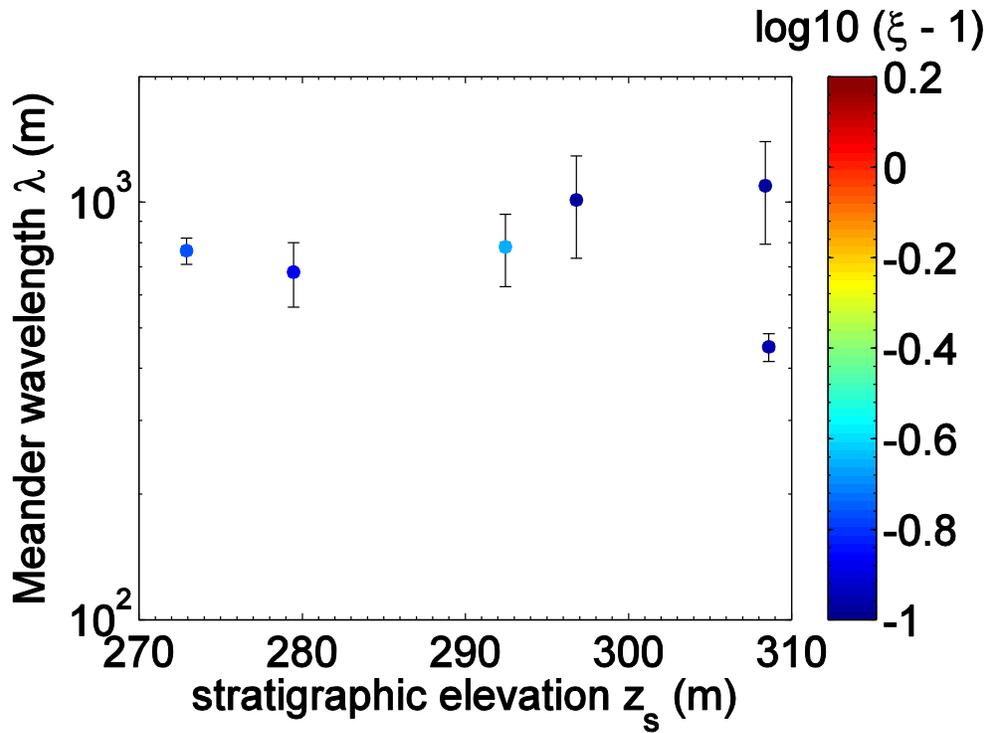

c)

**Figure S3.** λ versus $z_s$. Color corresponds to sinuosity ξ. **a)** Transect 1; **b)** Transect 2; **c)** Transect 3. The stratigraphic RMS error is 20 m for Transect 1; 5 m for Transect 2; and 67 m for Transect 3.

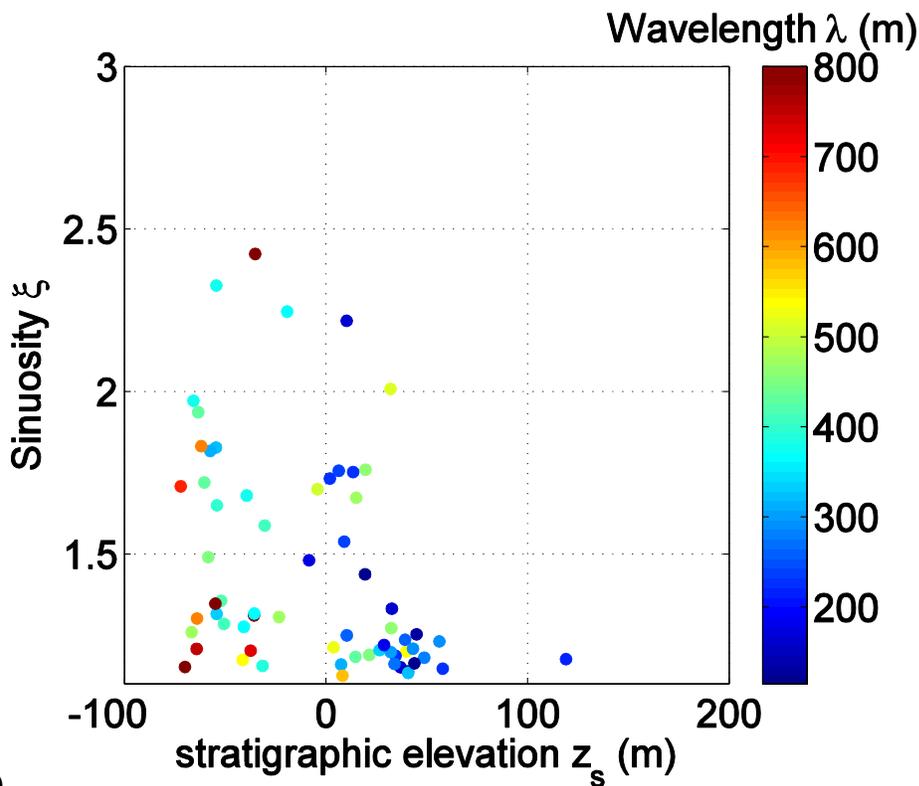

a)

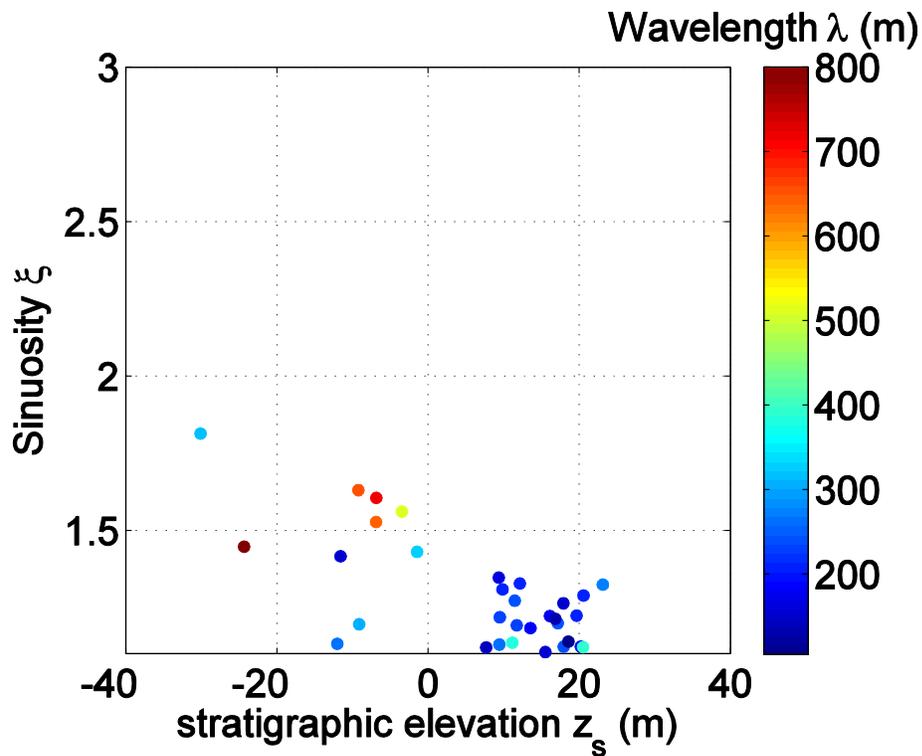

b)

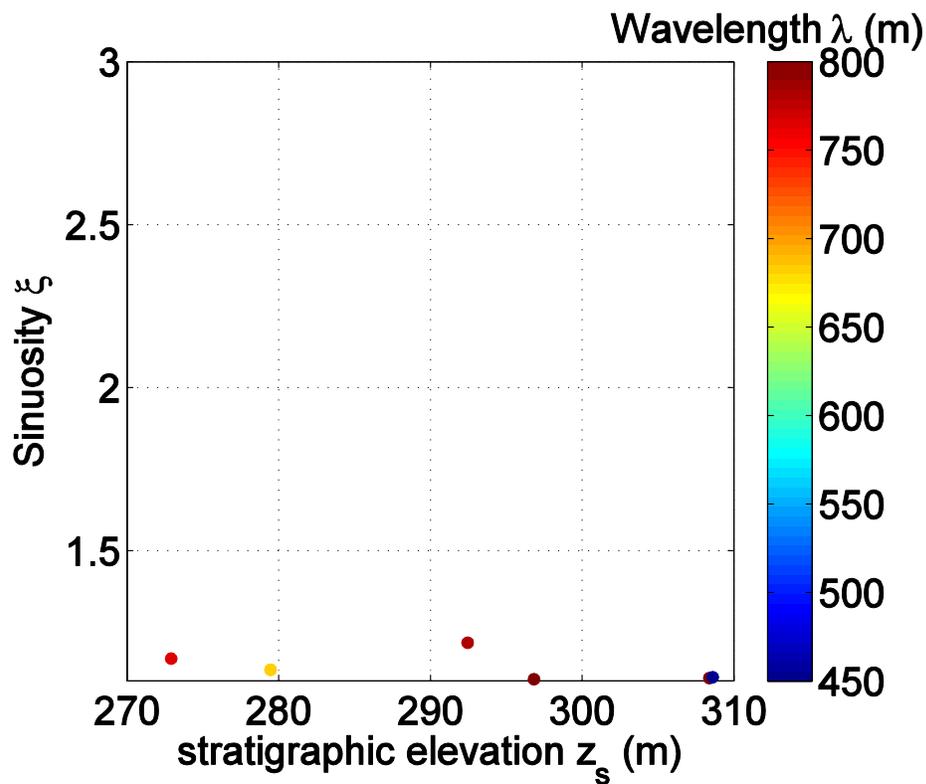

c)

**Figure S4.** ξ versus $z_s$. **a)** Transect 1; **b)** Transect 2; **c)** Transect 3. The stratigraphic RMS error is 20 m for Transect 1; 5 m for Transect 2; and 67 m for Transect 3.

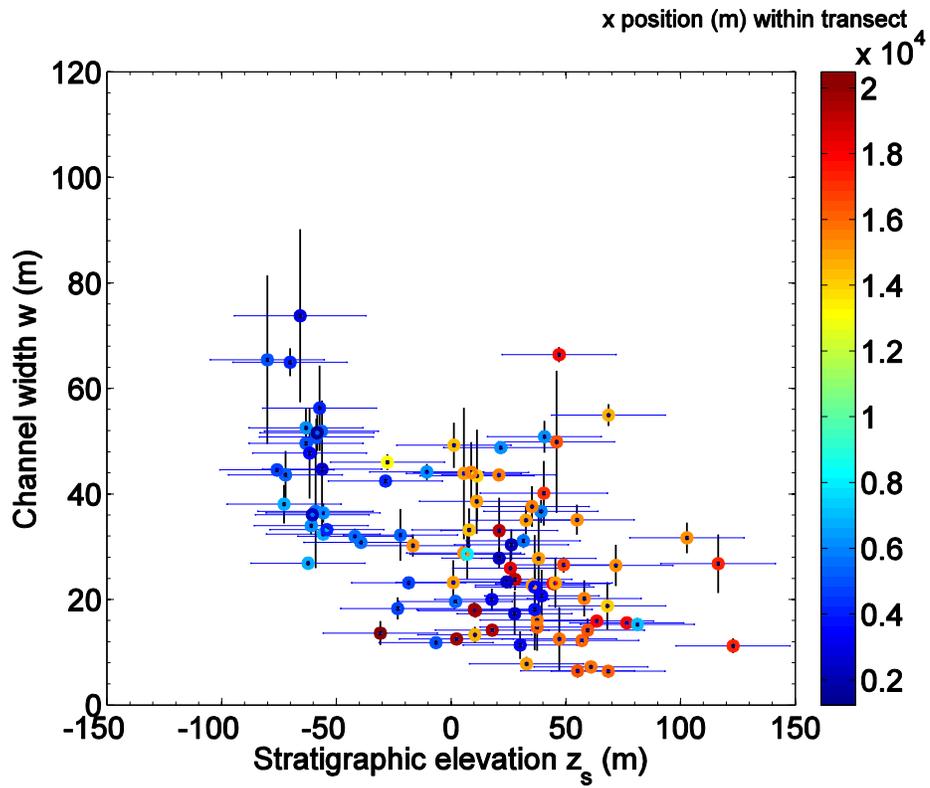

a)

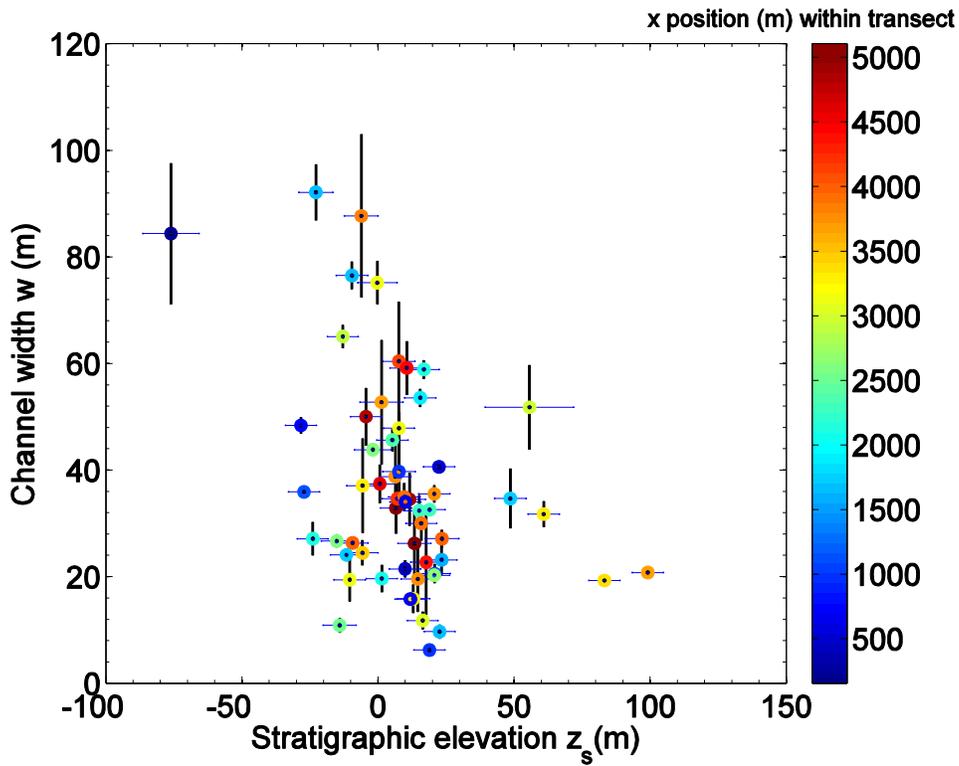

b)

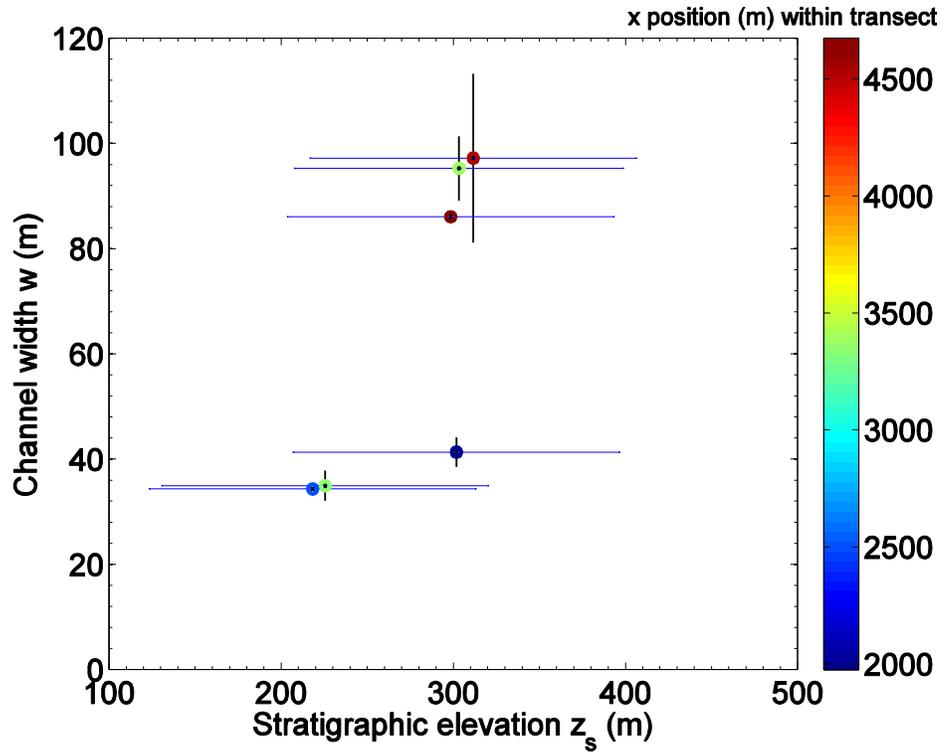

c)

**Figure S5.** *w* versus $z_s$ (with *w* collated by TieNumber). **a)** Transect 1; **b)** Transect 2; **c)** Transect 3. The stratigraphic RMS error is 20 m for Transect 1; 5 m for Transect 2; and 67 m for Transect 3.

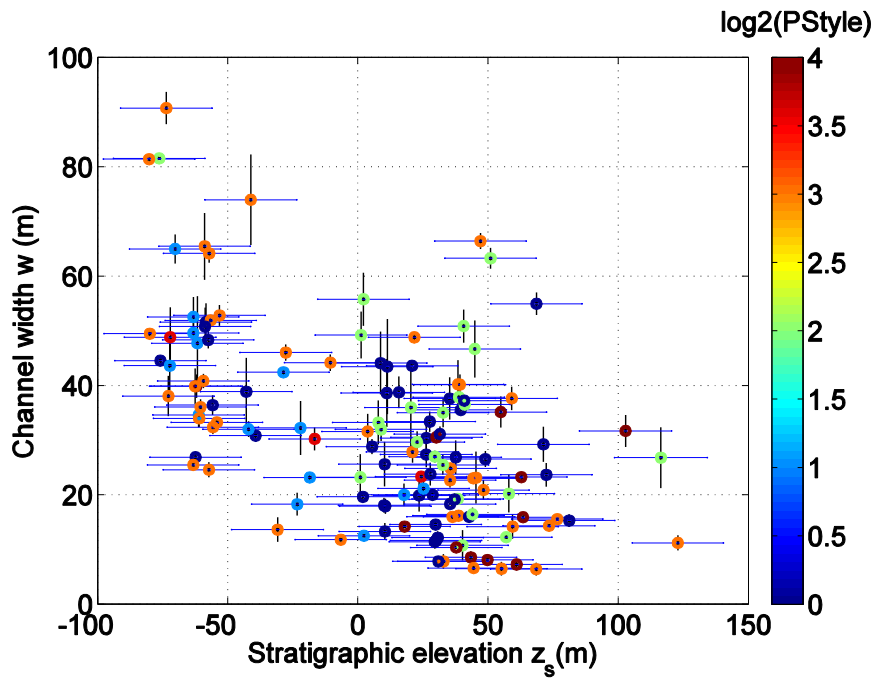

a)

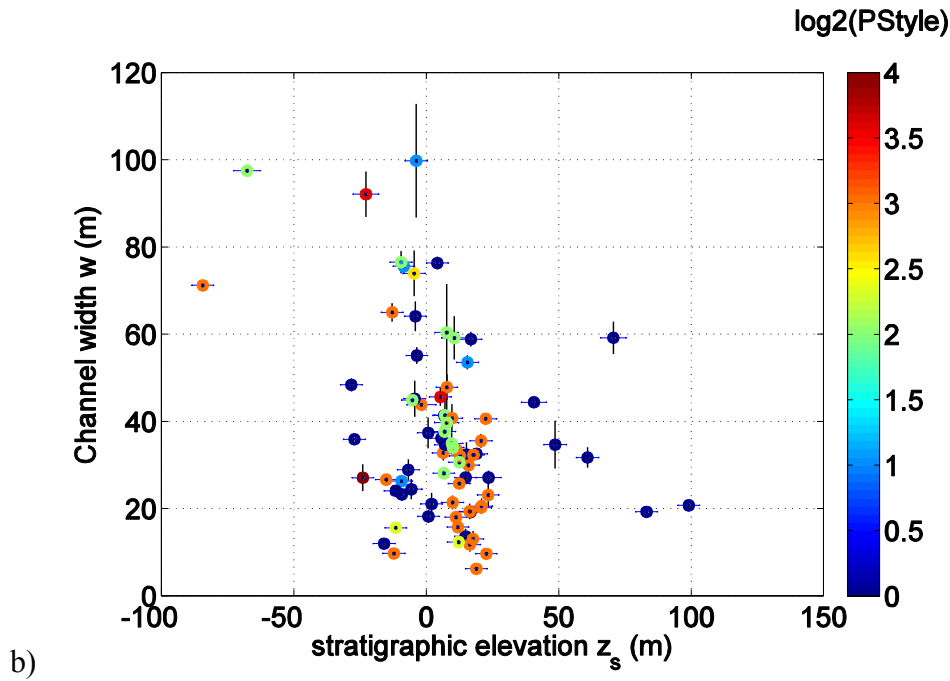

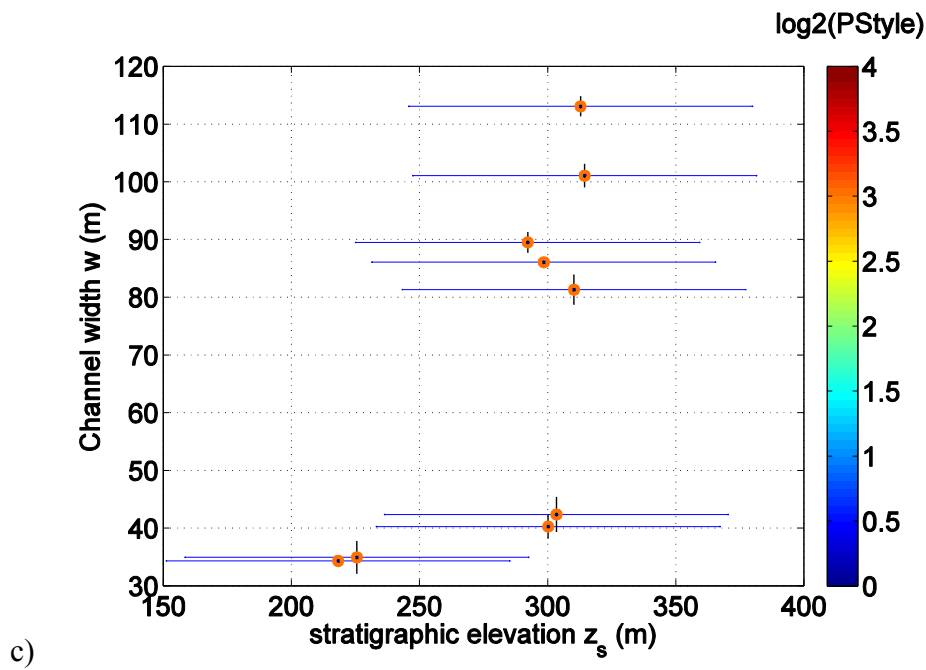

**Figure S6.** $w$ versus $z_s$ (individual $w$ bank-pair measurements). Colors correspond to preservation style:- dark blue (PStyle=1) for valleys (negative relief), light blue (PStyle=2) for part of clear meander/sinuous shape, green (PStyle=4) for negative-relief valleys containing a positive-relief deposit, orange (PStyle=8) for inverted channels, and red (PStyle=16) for double ridges only. **a)** Transect 1; **b)** Transect 2; **c)** Transect 3. The stratigraphic RMS error is 20 m for Transect 1; 5 m for Transect 2; and 67 m for Transect 3.

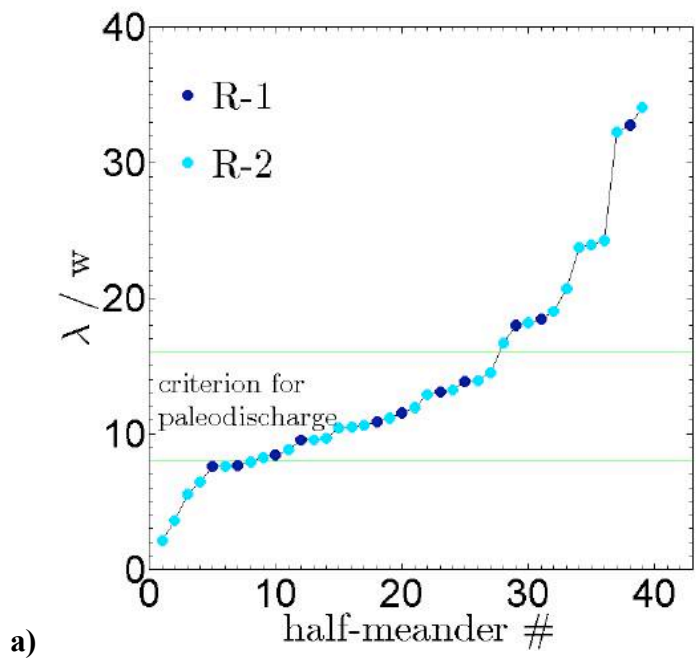

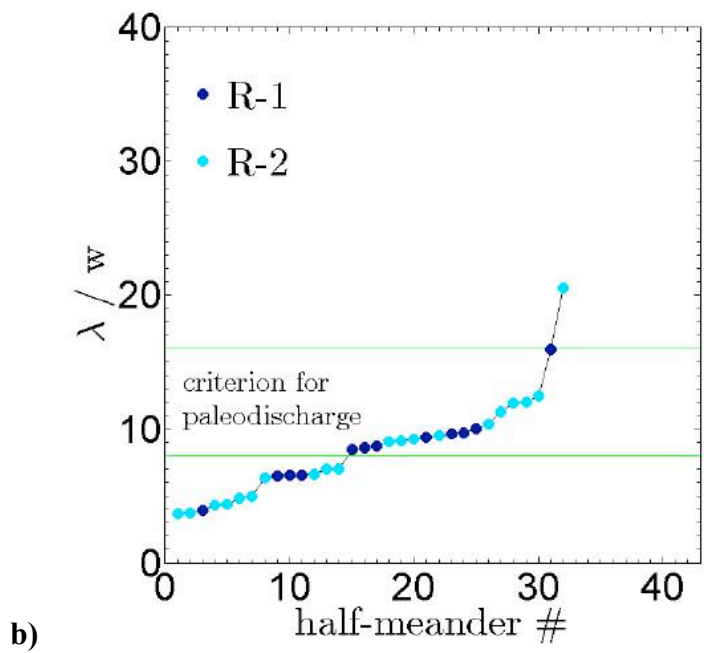

**Figure S7.** Results of crosscheck for cases where width and wavelength measurements were collocated. "Criterion for paleodischarge" is from Burr et al. (2010). **a)** Transect 1; **b)** Transect 2.